\documentclass[aip, jcp, notitlepage, floatfix, reprint,
  citeautoscript, twocolumn]{revtex4-1}

\usepackage{amsmath}
\usepackage{amsfonts,amssymb}
\usepackage{graphicx}
\usepackage{dcolumn}
\usepackage{miller}
\usepackage[table]{xcolor}
\usepackage[version=3]{mhchem}
\usepackage{transparent}
\usepackage{gensymb}
\usepackage{tabularx}
\usepackage{etoolbox}
\usepackage[hidelinks]{hyperref}
\usepackage{bm}
\usepackage{relsize}
\usepackage[normalem]{ulem}

\def\maxfloatwidth{%
  \ifdim\columnwidth>246.0pt
  300.0pt  \else
  \columnwidth
  \fi
}

\newcommand{\tbf}[1]{\textbf{#1}}

\newcommand{\mbf}[1]{\mathbf{#1}}
\newcommand{\mrm}[1]{\mathrm{#1}}
\newcommand{\tcr}[1]{\textcolor{black}{#1}}
\newcommand{\tcb}[1]{\textcolor{black}{#1}}
\newcommand{\etal}{\emph{et al.}}
\newcommand{\me}{\mathrm{e}}

\begin{document}

\title{Can molecular simulations reliably compare homogeneous and
  heterogeneous ice nucleation?}

\author{Dominic Atherton}

\author{Angelos Michaelides} 

\author{Stephen J. Cox} 
\affiliation{Yusuf Hamied Department of Chemistry, University of
  Cambridge, Lensfield Road, Cambridge CB2 1EW, United Kingdom}
\email{sjc236@cam.ac.uk}

\date{\today}

\begin{abstract}
  In principle, the answer to the posed titular question is
  undoubtedly `yes.' But in practice, requisite reference data for
  homogeneous systems have been obtained with a treatment of
  intermolecular interactions that is different from that typically
  employed for heterogeneous systems. In this article, we assess the
  impact of the choice of truncation scheme when comparing water in
  homogeneous and inhomogeneous environments. Specifically, we use
  explicit free energy calculations and a simple mean field analysis
  to demonstrate that using the `cut-and-shift' version of the
  Lennard-Jones potential (common to most simple point charge models
  of water) results in a systematic increase in the melting
  temperature of ice I$_{\rm h}$. In addition, by drawing an analogy
  between a change in cutoff and a change in pressure, we use existing
  literature data for homogeneous ice nucleation at negative pressures
  to suggest that enhancements due to heterogeneous nucleation may
  have been overestimated by several orders of magnitude.
\end{abstract}

\maketitle

\section{Introduction}
\label{sec:intro}

The formation of ice is a process of great importance across a broad
range of fields, from climate science
\cite{tan2016observational,slater2016blue} to biology
\cite{bar2016ice}. Obtaining a detailed molecular-level understanding
of both homogeneous nucleation (i.e., in the absence of foreign bodies
such as mineral particles) and heterogeneous nucleation (i.e., in the
presence of surfaces due to foreign bodies) has attracted major
research efforts from both experimental and simulation groups
\cite{murray2012ice,sosso2016crystal}. With regard to the latter, in a
bid to reduce computational cost, most molecular simulations employ
empirical potentials that approximately describe the interactions
between water molecules. While many types of empirical potentials
exist
\cite{molinero2009water,vega2011simulating,cisneros2016modeling},
simple point charge (SPC) models are one of the most commonly
used. Note that we use `SPC model' to refer to the general class of
water model detailed in Sec.~\ref{subsec:formulating} rather than the
specific water model of Ref.~\onlinecite{OrigSPC}. In addition to
being relatively simple and computationally efficient, an appealing
feature of SPC models is that they preserve the donor-acceptor nature
of water's hydrogen-bond network, which can be especially important
for heterogeneous nucleation, e.g., in the presence of kaolinite
\cite{cox2013microscopic,sosso2016microscopic}.

\begin{figure}[tb]
  \includegraphics[width=8cm]{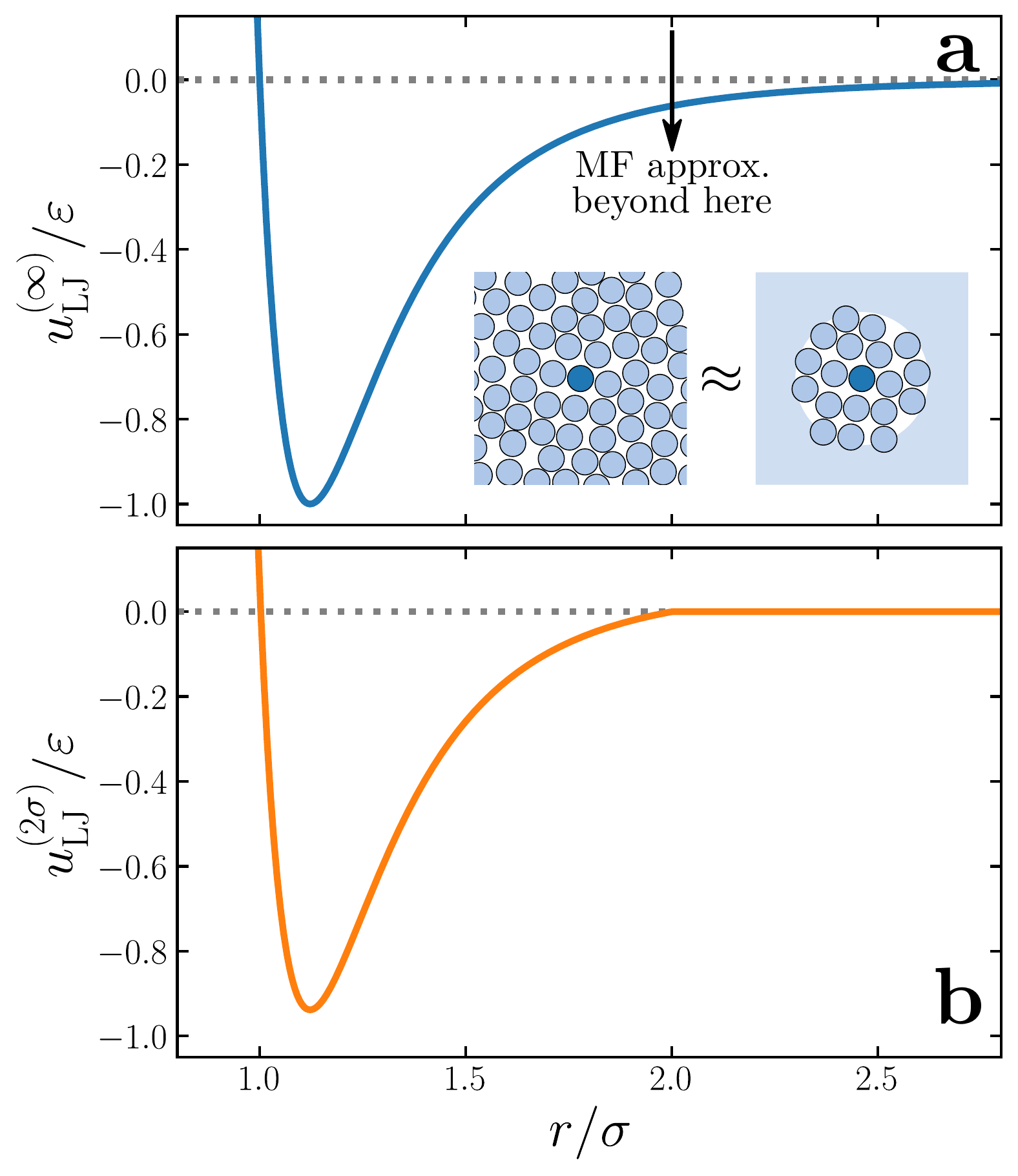}
  \caption{The two variations of the LJ potential studied in this
    article. (a) The solid blue curve shows $u_{\rm LJ}^{(\infty)}$
    given by Eq.~\ref{eqn:LJ}. For \emph{homogeneous} systems
    $u^{(\infty)}_{\rm LJ}$ is well-approximated by truncating
    interactions at a given cutoff (indicated by the arrow), and
    applying `tail corrections' to account for neglected interactions;
    this is equivalent to a mean field approximation. Inset: schematic
    representation of the tail correction procedure. The blue shaded
    region indicates that interactions between the tagged particle
    (dark blue) and those beyond the cutoff radius are accounted for
    in an average sense. (b) The `cut-and-shift' variant of the LJ
    potential (see Eq.~\ref{eqn:ulj_ts}) has vanishing interactions
    beyond the cutoff. It is a different potential with different
    properties compared to $u_{\rm LJ}^{(\infty)}$. In both examples,
    the cutoff radius is $2\sigma$.}
  \label{fig:intro}
\end{figure}

To ensure short-ranged repulsion between molecules, most commonly used
SPC water models, at least formally, employ the Lennard-Jones (LJ)
potential \cite{lj1931cohesion},
\begin{equation}
  \label{eqn:LJ}
  u^{(\infty)}_{\rm LJ}(r) = 4\varepsilon\Bigg[\bigg(\frac{\sigma}{r}\bigg)^{12}-\bigg(\frac{\sigma}{r}\bigg)^{6}\Bigg],
\end{equation}
which is parameterized by an energy scale $\varepsilon$ and length
scale $\sigma$, and where $r$ indicates the distance between two water
molecules (usually the separation between their oxygen
atoms). Figure~\ref{fig:intro}a shows $u^{(\infty)}_{\rm LJ}$.

In addition to explicit electrostatic interactions between water
molecules, the $-(\sigma/r)^6$ term contributes to the cohesive energy
of the system. Despite being the basis for most SPC water models,
however, $u^{(\infty)}_{\rm LJ}$ is rarely sampled explicitly; due to
the infinite range of the attractive $-(\sigma/r)^6$ term, it is
common to truncate $u^{(\infty)}_{\rm LJ}$ in some fashion (see, e.g.,
Refs.~\onlinecite{FrenkelSmit2002sjc,allen1987computer}). Two common
procedures, which we detail in Sec.~\ref{subsec:formulating}, are to
use `tail corrections' or to `cut-and-shift', as shown in
Figs.~\ref{fig:intro}a and~\ref{fig:intro}b, respectively. By
comparing these plots it can be seen that, while similar, these two
truncation procedures result in different intermolecular potentials,
and will in general have different properties. For example,
thermodynamic properties such as interfacial tension and the location
of phase boundaries are known to be affected
\cite{smit1992phase,johnson1993lennard,baidakov2000effect,hafskjold2019thermodynamic,ghoufi2016computer,fitzner2017communication}.

Why then, would another article that investigates the effects of
truncating the LJ potential be useful? Put simply, the phase behavior
of SPC water models has been studied extensively using the tail
corrected truncation scheme
\cite{sanz2004phase,vega2005melting,abascal2005general,abascal2005potential}. And
the same can be said for most calculations of homogeneous ice
nucleation rates
\cite{espinosa2014homogeneous,haji2015direct,espinosa2016interfacial,bianco2021anomalous}. Yet,
as we will discuss in more detail below, the use of tail corrections
makes direct comparison to inhomogeneous systems challenging. While a
simple approach to mitigate discrepancies between homogeneous and
inhomogeneous systems would be consistent use of cut-and-shift
potentials, it is unreasonable to expect that each study of
heterogeneous nucleation is accompanied by: (i) a full recalculation
of the melting temperature or phase diagram; and (ii) recomputation of
the homogeneous nucleation rate. [To give a sense of perspective, in
  Ref.~\onlinecite{haji2015direct} over $21\times 10^6$ CPU hours were
  required to compute the homogeneous nucleation rate with forward
  flux sampling (FFS).] In this article, we address the first issue
directly, by outlining a procedure to approximately predict the change
in melting temperature between the tail-corrected and cut-and-shift
systems. We then combine our results with those in
Ref.~\onlinecite{bianco2021anomalous} to estimate the impact on the
comparison of homogeneous and heterogeneous nucleation rates.

\subsection{Formulating the problem}
\label{subsec:formulating}

We now detail the tail-correction and cut-and-shift truncation
schemes, as well as illustrate the inconsistencies that appear between
homogeneous and inhomogeneous systems. We are concerned with SPC water
models that formally have a potential energy function of the kind
\begin{equation}
  \label{eqn:Utot}
  U^{(\infty)}(\mbf{R}^N) = \sum_{i<j}^Nu^{(\infty)}_{\rm LJ}(|\mbf{r}^{\rm (O)}_{ij}|)
  + \sum_{i<j}^N\sum_{\alpha,\beta}\frac{q^{(\alpha)}q^{(\beta)}}{|\mbf{r}^{(\alpha)}_{j}-\mbf{r}^{(\beta)}_{i}|},
\end{equation}
where $\mbf{R}^N$ denotes the set of atomic positions for a
configuration of $N$ water molecules, $\mbf{r}_{ij}^{\rm (O)}$ is the
separation vector between the oxygen atoms of molecules $i$ and $j$,
and $q^{(\alpha)}_i$ is the charge of site $\alpha$, located at
$\mbf{r}_{i}^{(\alpha)}$, of molecule $i$. (We adopt a unit system in
which $4\pi\epsilon_0 = 1$, where $\epsilon_0$ is the permittivity of
free space.) The second set of sums in Eq.~\ref{eqn:Utot}, which we
will denote $U_{\rm elec}$ hereafter, describes electrostatic
interactions between molecules, while the first set of sums involve
the LJ potential. For SPC models of water, the choice of
$u^{(\infty)}_{\rm LJ}$ is rooted in grounds of convention and
convenience, rather than having any deep theoretical
justification. Nonetheless, SPC models of the kind formally described
by Eq.~\ref{eqn:Utot} have been, are, and will likely continue (at
least in the near future) to be the foundation for many molecular
simulations of water's condensed phases.

\begin{figure}[tb]
  \includegraphics[width=8cm]{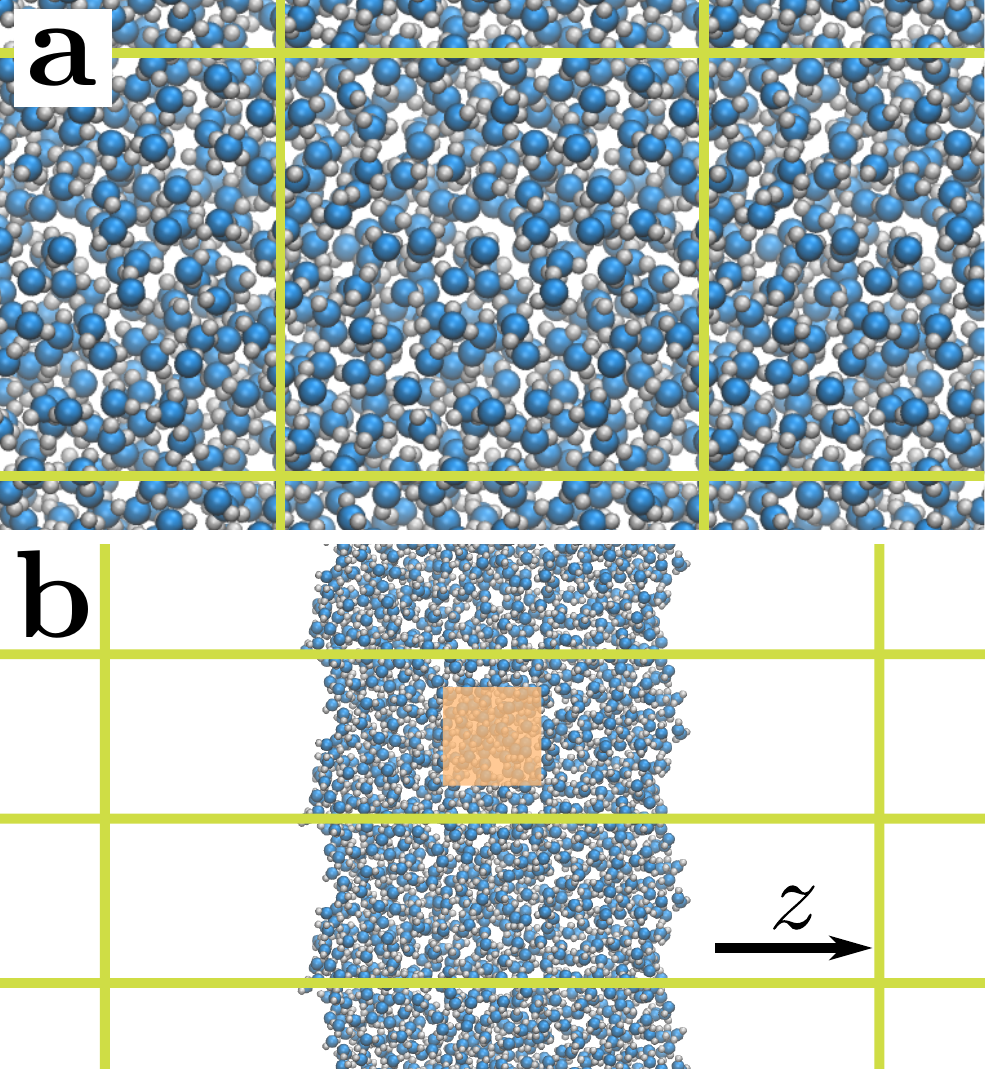}
  \caption{Typical simulation geometries for water. (a) Homogeneous
    bulk water simulated under 3D bulk periodic boundary conditions
    such that the average density is independent of position,
    $\langle\rho(\mbf{r})\rangle = \bar{\rho}(p)$. (b) The slab
    geometry employed to study interfacial systems also employs
    periodic boundary conditions (often in 3D) and has a density
    profile that varies with position $z$ along the surface normal,
    such that in general $\langle\rho(z)\rangle\neq\bar{\rho}$. For
    thick enough slabs, regions far removed from the interfaces (as
    indicated, e.g., by the orange box)) have an average density equal
    to $\bar{\rho}(p=0)$ for the homogeneous system. Green lines
    indicate the boundaries of the periodically repeated simulation
    cells.}
  \label{fig:Schematic}
\end{figure}

So far, we have referred to SPC models of water that are `formally'
described by the potential energy given by Eq.~\ref{eqn:Utot}. But as
already mentioned, in practice $u^{(\infty)}_{\rm LJ}$ is usually
truncated in some fashion
\cite{FrenkelSmit2002sjc,allen1987computer}. For the tail-correction
scheme, one employs a simple truncation,
\begin{equation}
  \label{eqn:uLJsimplecut}
  u_{\rm LJ}^{(r_{\rm c}\to\infty)}(r) =
  \begin{cases}
    u^{(\infty)}_{\rm LJ}(r), & r\le r_{\rm c}, \\
    0, & r>r_{\rm c},
  \end{cases}
\end{equation}
and then approximately accounts for the effects of truncation by
adding a mean field (MF) correction,
\begin{equation}
  \label{eqn:DelMF-U}
  \frac{\Delta_{\rm MF}U(r_{\rm c})}{N} =
  \frac{8\pi\epsilon\bar{\rho}\sigma^{3}}{9}\Bigg[\bigg(\frac{\sigma}{r_{\rm c}}\bigg)^9-3\bigg(\frac{\sigma}{r_{\rm c}}\bigg)^3\Bigg],
\end{equation}
to the total potential energy:
\begin{align}
  \label{eqn:Utot-MF}
  &U^{(\infty)}(\mbf{R}^N) \approx U^{(r_{\rm c}\to\infty)}(\mbf{R}^N) \nonumber \\
  &= \sum_{i<j}^Nu^{(r_{\rm c}\to\infty)}_{\rm LJ}(|\mbf{r}^{\rm (O)}_{ij}|) + \Delta_{\rm MF}U(r_{\rm c})
  + U_{\rm elec}(\mbf{R}^N),
\end{align}
where $\bar{\rho}$ is the average number density. The superscript
`$(r_{\rm c}\to\infty)$' indicates that, when used in combination with
$\Delta_{\rm MF}U(r_{\rm c})$, a system that employs $u_{\rm
  LJ}^{(r_{\rm c}\to\infty)}$ satisfies $U^{(r_{\rm c}\to\infty)}
\approx U^{(\infty)}$; this is reasonable provided that $g_{\rm
  OO}(r\ge r_{\rm c})\approx 1$, where $g_{\rm OO}$ is the
oxygen-oxygen pair correlation function. In a similar spirit, the
pressure can also be corrected in a MF fashion,
\begin{equation}
  \label{eqn:DelMF-P}
  \Delta_{\rm MF}p(r_{\rm c})
  = \frac{32\pi\epsilon\bar{\rho}^2\sigma^3}{9}\Bigg[\bigg(\frac{\sigma}{r_{\rm c}}\bigg)^9-\frac{3}{2}\bigg(\frac{\sigma}{r_{\rm c}}\bigg)^3\Bigg].
\end{equation}
A comment is in order concerning the functional form of $u_{\rm
  LJ}^{(r_{\rm c}\to\infty)}$ (Eq.~\ref{eqn:uLJsimplecut}). The
discontinuity at $r_{\rm c}$ suggests the presence of impulsive
forces. \tcb{Impulsive forces, however, are challenging to implement
  in molecular dynamics simulations, and it is standard practice to
  neglect them entirely. Moreover, as interactions} beyond $r_{\rm c}$
are not neglected in $U^{(r_{\rm c}\to\infty)}$, but instead accounted
for in a mean field fashion, \tcb{we argue (see SM) that this neglect
  of impulsive forces is in fact consistent with the use of
  $\Delta_{\rm MF}U$ and $\Delta_{\rm MF}p$, as it accounts for a
  pointwise cancellation of impulsive forces.}

The alternative cut-and-shift truncation scheme is:
\begin{equation}
  \label{eqn:ulj_ts}
  u_{\rm LJ}^{(r_{\rm c})}(r) =
  \begin{cases}
    u^{(\infty)}_{\rm LJ}(r) - u^{(\infty)}_{\rm LJ}(r_{\rm c}), & r\le r_{\rm c}, \\
    0, & r>r_{\rm c},
  \end{cases}
\end{equation}
such that the total potential energy is
\begin{equation}
  \label{eqn:Utot-cs}
  U^{(r_{\rm c})}(\mbf{R}^N) = \sum_{i<j}^Nu^{(r_{\rm c})}_{\rm LJ}(|\mbf{r}^{\rm (O)}_{ij}|)
  + U_{\rm elec}(\mbf{R}^N).
\end{equation}
We will use the superscript `$(r_{\rm c})$' to indicate that $u_{\rm
  LJ}^{(r_{\rm c})}$ is used. (We will, on occasion, drop the
superscript notation, either because it is clear from context which
truncation scheme is relevant, or because it is unimportant to
differentiate between truncation schemes. When a numerical value of
$r_{\rm c}$ is specified, it will be given in \AA ngstrom, though we
will omit units from the superscript.) For simulations of systems in
the canonical ($NVT$) ensemble, dynamics are unaffected by the choice
of $U^{(r_{\rm c})}$ vs. $U^{(r_{\rm c}\to\infty)}$. The pressure,
however, is sensitive to the choice of truncation scheme:
\begin{equation}
  \label{eqn:Pdiff}
  p^{(r_{\rm c}\to\infty)} \approx p^{(r_{\rm c})}+\Delta_{\rm MF}p(r_{\rm c}).
\end{equation}
The implication of Eq.~\ref{eqn:Pdiff} is that dynamics in the
isothermal-isobaric ($NpT$) ensemble are affected by the choice of
$U^{(r_{\rm c})}$ vs. $U^{(r_{\rm c}\to\infty)}$. Furthermore, systems
employing $U^{(r_{\rm c}\to\infty)}$ and $U^{(r_{\rm c})}$ will have,
for the same $r_{\rm c}$, different equations of state
\cite{smit1992phase,johnson1993lennard}.

\begin{figure}[tb]
  \centering
  \includegraphics[width=8cm]{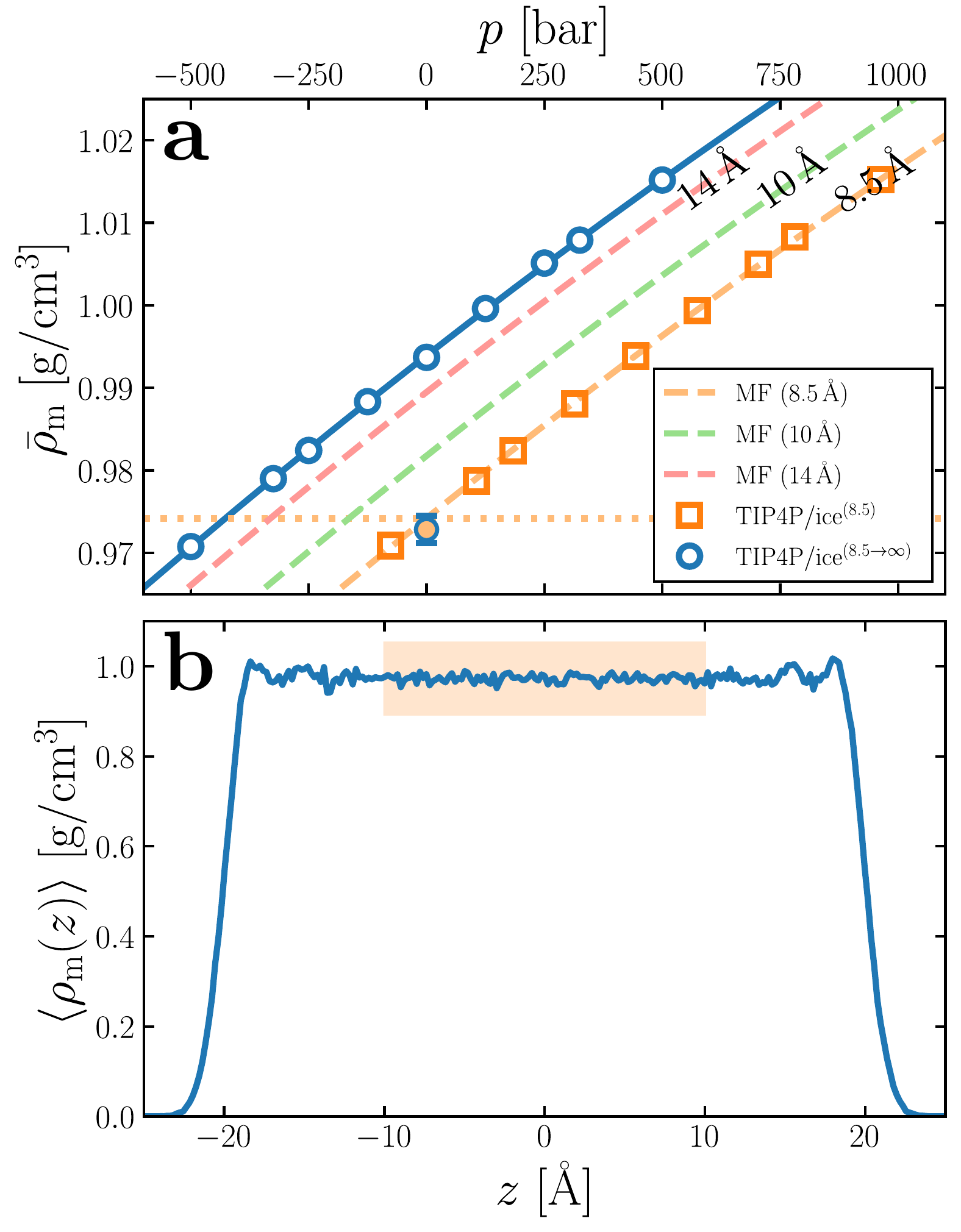}
  \caption{Evaluating the impact of $r_{\rm c}$ on $\bar{\rho}_{\rm
      m}$ for liquid TIP4P/ice at 300\,K. (a) $\bar{\rho}_{\rm m}(p)$
    for a homogeneous system. White-filled circles show results from
    constant-$p$ simulations of TIP4P/ice$^{(8.5\to\infty)}$, and the
    solid blue line indicates a quadratic fit. Dashed lines indicate
    MF predictions (Eq.~\ref{eqn:Pdiff}, see also
    Sec.~\ref{sec:Tm-MF}) for different $r_{\rm c}$, as indicated in
    the legend (the lines are also labeled). Orange squares show
    results from constant-$p$ simulations of TIP4P/ice$^{(8.5)}$. The
    dotted line indicates $\bar{\rho}_{\rm m}(p=0)$ for
    TIP4P/ice$^{(8.5)}$, which intercepts the
    TIP4P/ice$^{(8.5\to\infty)}$ results at $p\approx -427$\,bar. (b)
    $\langle\rho_{\rm m}(z)\rangle$ for a film of
    TIP4P/ice$^{(8.5\to\infty)}$ in contact with its vapor (only part
    of the simulation cell is shown). Spatially averaging
    $\langle\rho_{\rm m}(z)\rangle$ in the slab's interior, as
    indicated by the shaded region, gives an estimate for
    $\bar{\rho}_{\rm m}(0)$, which is plotted with the orange-filled
    circle in (a).}
  \label{fig:RhoVsPress_300K}
\end{figure}

Implicit in our above discussion of MF corrections is that the system
is homogeneous, such that the average equilibrium density
$\langle\rho(\mbf{r})\rangle = \bar{\rho}$ does not depend upon the
position $\mbf{r}$ in the fluid, as shown in
Fig.~\ref{fig:Schematic}a. If the system of interest is inhomogeneous,
such as liquid water in coexistence with its vapor, a typical
simulation approach is to employ the $NVT$ ensemble with a cuboidal
cell that has an elongated dimension along the average surface normal;
such a scenario is depicted in Fig.~\ref{fig:Schematic}b. As
$\Delta_{\rm MF}U$ and $\Delta_{\rm MF}p$ do not affect dynamics in
the $NVT$ ensemble, effects of using $U^{(r_{\rm c}\to\infty)}$ for
inhomogeneous systems would perhaps seem benign, resulting simply in a
shift of the energy, i.e., $U^{(r_{\rm
    c}\to\infty)}(\mbf{R}^N)-U^{(\infty)}(\mbf{R}^N)=\text{const.}$

Potential problems arise, however, concerning thermodynamic
consistency between the homogeneous and inhomogeneous systems. This is
demonstrated in Fig.~\ref{fig:RhoVsPress_300K} for
TIP4P/ice\cite{abascal2005potential}---a commonly used SPC water model
for studying ice formation---at 300\,K, with $r_{\rm c}=8.5$\,\AA.
Fig.~\ref{fig:RhoVsPress_300K}a shows the average mass density
$\bar{\rho}_{\rm m}(p)$ obtained from simulations of the homogeneous
fluid employing either TIP4P/ice$^{(8.5\to\infty)}$ or
TIP4P/ice$^{(8.5)}$. Fig.~\ref{fig:RhoVsPress_300K}b shows the
equilibrium mass density profile $\langle\rho_{\rm m}(z)\rangle$ for a
film of TIP4P/ice$^{(8.5\to\infty)}$ water approximately 40\,\AA{}
thick, with its liquid/vapor interface spanning the $xy$
plane.\footnote{Figure~\ref{fig:RhoVsPress_300K}b in fact shows the
  number density profile converted to mass density. While the two
  profiles will differ slightly near the interface, they are the same
  in the bulk region of interest in this study.} Owing to the low
vapor pressure of water, $p\approx 0$ in the vapor phase. As the
normal component of the pressure tensor is independent of $z$ for a
planar interface, and furthermore isotropic for $z$ in a bulk-like
fluid region, it immediately follows that $p\approx 0$ deep in the
slab's interior \cite{rowlinson2002molecular}. Thermodynamic
consistency then requires that
\[
\frac{1}{\ell_{\rm bulk}}\int_{\ell_{\rm bulk}}\!\mrm{d}z\,\langle\rho_{\rm m}(z)\rangle = \bar{\rho}_{\rm m}(p=0)
\]
for TIP4P/ice$^{(8.5\to\infty)}$, where $\ell_{\rm bulk}$ is a length
over which $\langle\rho_{\rm m}(z)\rangle$ is bulk-like, as indicated,
e.g., by the orange rectangles in Figs.~\ref{fig:Schematic}b and
\ref{fig:RhoVsPress_300K}b. The result of such an averaging procedure
is indicated by the orange-filled circle in
Fig.~\ref{fig:RhoVsPress_300K}a; it is clearly inconsistent with
$\bar{\rho}_{\rm m}(p=0)$ obtained from the homogeneous
TIP4P/ice$^{(8.5\to\infty)}$ simulation. As dynamics in the $NVT$
ensemble are unaffected by the choice of $U^{(r_{\rm c})}$
vs. $U^{(r_{\rm c}\to\infty)}$ we might expect, and indeed observe,
that the result is instead consistent with $\bar{\rho}_{\rm m}(p=0)$
for TIP4P/ice$^{(8.5)}$.

While schemes for effectively sampling $U^{(\infty)}$ do exist for
heterogeneous systems (e.g., one can treat the attractive
$-(\sigma/r)^6$ term in a Ewald
fashion\cite{in2007application,alejandre2010surface,lopez2002thermodynamic,lopez2003simulation},
or use mean-field corrections that take the heterogeneous nature of
the system into
account\cite{miguez2013influence,janecek2006long,salomons1991atomistic,guo1997long1,guo1997long2,de2012semi})
their use is relatively limited compared to that of SPC water models.
And, as discussed in Sec.~\ref{sec:intro}, no discrepancies would be
observed with consistent use of $U^{(r_{\rm c})}$ for both the
homogeneous and inhomogeneous systems, but information concerning
phase behavior and homogeneous nucleation rates relevant to
$U^{(r_{\rm c})}$ systems is scarce. In
Sec.~\ref{sec:meltingEinstein}, we therefore assess the effect of
using $U^{(r_{\rm c})}$ instead of $U^{(r_{\rm c}\to\infty)}$ on the
melting temperature $T_{\rm m}$ of ice I$_{\rm h}$ for SPC models of
water. In particular we focus on TIP4P/ice \cite{abascal2005potential}
and TIP4P/2005 \cite{abascal2005general}, as these are most commonly
used in simulations of ice nucleation. We stress, however, that the
findings presented in this work readily extend to any SPC water model
of the kind formally described by Eq.~\ref{eqn:Utot}. To illustrate
our findings, we will focus exclusively on results for TIP4P/ice in
the main article, with results for TIP4P/2005 instead given in the
Supplementary Material (SM). In Sec.~\ref{sec:Nucleation} we then
estimate the impact of our findings on the comparison of homogeneous
and heterogeneous ice nucleation rates.

As an aside, before proceeding to discuss our main results, we mention
that our initial motivation for this study stemmed from recent work by
Wang \etal \cite{wang2020lennard}, who developed a new potential that
gives broadly similar behavior to the LJ potential, but does not
suffer, by construction, from ambiguities arising from the choice of
truncation scheme. While our preliminary investigations suggested that
the approach of Wang \etal{} could be used to develop workable SPC
models of water, we judged their performance insufficiently strong to
warrant introducing another set of SPC water models to the
community. We therefore adopt a more pragmatic approach in this
article by instead providing results and insights relevant to existing
SPC water models that are heavily used by practitioners of molecular
simulations.

\section{The melting point of ice I$_{\rm h}$ from free energy calculations}
\label{sec:meltingEinstein}

The central quantity under investigation in this study is the melting
point of ice I$_{\rm h}$ under conditions of vanishing pressure,
$p=0$\,bar. To obtain estimates of $T_{\rm m}$ for TIP4P/ice$^{(r_{\rm
    c})}$ and TIP4P/2005$^{(r_{\rm c})}$, we need to establish the
chemical potential at $p=0$\,bar for both the ice [$\mu_{\rm ice}(T)$]
and liquid [$\mu_{\rm liq}(T)$] phases: the point of intersection is
$T_{\rm m}$. To compute $\mu_{\rm ice}$, we will adopt the
Frenkel-Ladd approach \cite{frenkel1984new}, adapted by Vega and
co-workers for rigid SPC water models
\cite{noya2008computing,vega2008determination,aragones2013free}. As
this approach has been detailed elsewhere, we present a detailed
overview of our workflow in the SM, and discuss only the most salient
aspects of the methodology in the main article.

First, we equilibrate a crystal of ice I$_{\rm h}$ comprising 768
molecules at a temperature $T_{\rm i}$ to obtain the average cell
parameters.  The cell parameters are then fixed to their average
values, and the structure `minimized' by a low temperature simulation
at 0.1\,K. (We adopt this approach as the standard minimizers in
\texttt{LAMMPS} \cite{plimpton1995sjc} are incompatible with the
\texttt{RATTLE} algorithm \cite{andersen1983rattle} used to impose the
rigid body constraints of the water molecules.) Our reference
structure is then this minimized crystal structure with no
intermolecular interactions, and with the oxygen and hydrogen atoms of
each water molecule tethered to their positions by a harmonic
potential with spring constants $k_{\rm O}$ and $k_{\rm H}$,
respectively. The difference in Helmholtz free energy (per molecule)
$\Delta_{\rm r2i}a$ between this reference system and the interacting
ice crystal of interest is then calculated by thermodyamic integration
\cite{kirkwood1935statistical}, at temperature $T_{\rm i}$. The rigid
body constraints, however, mean that we do not know the free energy of
the reference system. We therefore define a `sub-reference' system
(with free energy $a_{\rm sub}$ that is calculated analytically) in
which only the oxygen atoms of the water molecules are tethered, and
compute the Helmholtz free energy between the sub-reference and
reference systems $\Delta_{\rm s2r}a$, also by thermodynamic
integration. The free energy of the ice crystal is then
\begin{equation}
  \label{eqn:aice}
  a_{\rm ice} = a_{\rm kin} + a_{\rm sub} + \Delta_{\rm s2r}a + \Delta_{\rm r2i}a - k_{\rm B}T_{\rm i}\ln\frac{3}{2} - k_{\rm B}T_{\rm i}\ln 2,
\end{equation}
with $a_{\rm kin} = 3k_{\rm B}T_{\rm i}\ln(T_0/T_{\rm i})$, where
$T_0$ is a reference temperature \tcr{(see SM)}. We use $T_0=272$\,K
throughout this article. The final two terms respectively account for
the Pauling entropy arising from proton disorder in ice I$_{\rm h}$,
and the fact that the reference system does not respect the
permutational invariance of the two protons in a water molecule
\cite{aragones2013free}. The chemical potential is, in general,
obtained from $\beta\mu_{\rm ice} = \beta a_{\rm ice} + \beta
p/\bar{\rho}$; as $p = 0$\,bar, we simply have $\beta\mu_{\rm ice} =
\beta a_{\rm ice}$. ($\beta = 1/k_{\rm B}T$, where $k_{\rm B}$ is
Boltzmann's constant.) We note that there have been extensive studies
to understand the effects of finite system size on the calculation of
free energies for solids (see Ref.~\onlinecite{vega2008determination}
for a detailed discussion). Previous simulation studies suggest that
the system size we use (768 molecules) is large enough to obtain a
reliable estimate of $T_{\rm m}$ for ice I$_{\rm
  h}$\cite{reinhardt2021quantum}. Moreover, it is likely that any
finite size effects will largely cancel when comparing the two
truncation schemes considered in this study.

For the liquid, we equilibrate a system comprising 360 molecules at
$T_{\rm i}$ to obtain an estimate of $\bar{\rho}$. At this density, we
then calculate the change in free energy $\Delta_{\rm LJ2w}a$ between
the LJ fluid and the SPC water model under investigation using
thermodynamic integration. For systems that employ $U^{(r_{\rm
    c}\to\infty)}$, we determine the excess free energy of the LJ
fluid $a^{(r_{\rm c}\to\infty)}_{\rm LJ,ex}\approx a^{(\infty)}_{\rm
  LJ,ex}$ from the equation of state. (For consistency with previous
calculations of water's phase diagram \cite{vega2008determination}, we
use the equation of state of Johnson \etal\cite{johnson1993lennard})
For systems using $U^{(r_{\rm c})}$, we must also compute the free
energy difference $\Delta_{\rm tc2cs}a$ between the $U^{(r_{\rm
    c}\to\infty)}$ and $U^{(r_{\rm c})}$ systems. The free energy of
the liquid is then
\begin{equation}
  \label{eqn:aliq}
  a_{\rm liq}^{(r_{\rm c})} = a_{\rm id} + a^{(r_{\rm c}\to\infty)}_{\rm LJ,ex} + \Delta_{\rm tc2cs}a + \Delta_{\rm LJ2w}a, 
\end{equation}
where $a_{\rm id} = k_{\rm B}T_{\rm i}\ln\big(\bar{\rho}(T_0/T_{\rm
  i})^3\big) - k_{\rm B}T_{\rm i}$ \tcr{(see SM)}. An analogous
expression holds for $a_{\rm liq}^{(r_{\rm c}\to\infty)}$, except that
$\Delta_{\rm tc2cs}a$ is omitted. The chemical potential is simply
$\beta\mu_{\rm liq}=\beta a_{\rm liq}$.

Once the chemical potential has been established at $T_{\rm i}$, we
establish its temperature dependence by integrating the
Gibbs-Helmholtz equation,
\begin{equation}
  \label{eqn:GibbsHelmholtz}
  \beta\mu_{\rm ice}(T) = \beta_{\rm i}\mu_{\rm ice}(T_{\rm i})
  - \int_{T_{\rm i}}^{T}\!\mrm{d}t\,\frac{h_{\rm ice}(t)}{k_{\rm B}t^2},
\end{equation}
where $h_{\rm ice}$ is the enthalpy per molecule of ice, and
$\beta_{\rm i} = 1/k_{\rm B}T_{\rm i}$. An analogous expression holds
for $\beta\mu_{\rm liq}$.

\begin{figure}[tb]
  \centering
  \includegraphics[width=8cm]{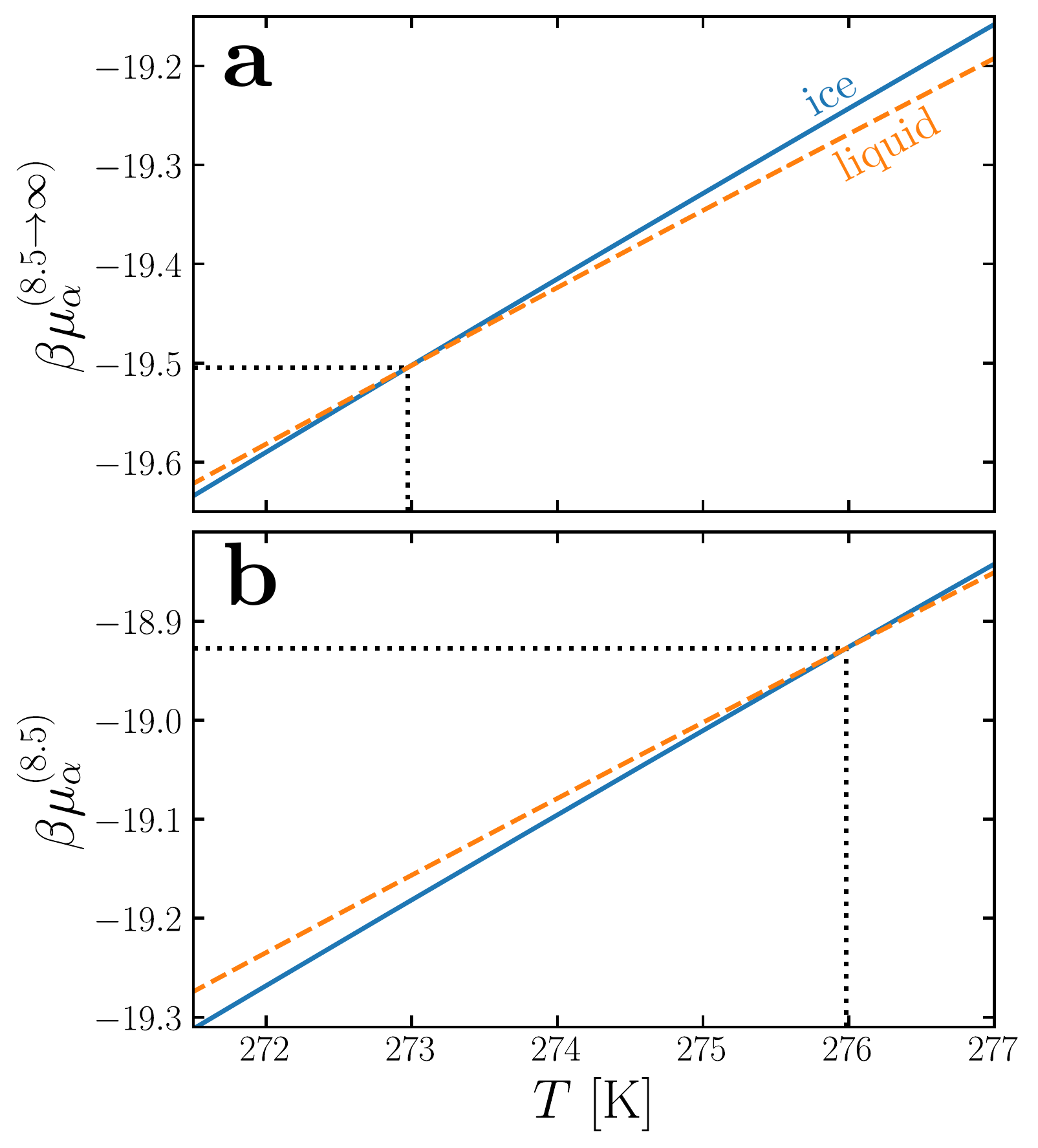}
  \caption{Locating coexistence: $\beta\mu_\alpha(T)$ at 0\,bar, with
    $\alpha = \text{`ice' or `liq'}$, for (a)
    TIP4P/ice$^{(8.5\to\infty)}$ and (b) TIP4P/ice$^{(8.5)}$. $T_{\rm
      m}$ is determined from the point of interception, as indicated
    by the black dotted lines, with $T^{(8.5\to\infty)}_{\rm m} =
    273.0$\,K and $T^{(8.5)}_{\rm m}=276.0$\,K.}
  \label{fig:BetaMuVsTemp_TailShift}
\end{figure}

In Fig.~\ref{fig:BetaMuVsTemp_TailShift}a, we present
$\beta\mu^{(8.5\to\infty)}_{\rm ice}(T)$ and
$\beta\mu^{(8.5\to\infty)}_{\rm liq}(T)$, from which we determine
$T^{(8.5\to\infty)}_{\rm m}\approx 273.0$\,K. This is in good
agreement with $T^{(8.5\to\infty)}_{\rm m} = 272\pm 6$\,K at
$p=1$\,bar obtained by Vega and co-workers
\cite{vega2005melting,abascal2005potential}. The results for
TIP4P/ice$^{\rm (8.5)}$ are shown in
Fig.~\ref{fig:BetaMuVsTemp_TailShift}b. It is clear that using
$U^{(r_{\rm c})}$ instead of $U^{(r_{\rm c}\to\infty)}$ results in an
apparent increase of the melting temperature, with $T^{(8.5)}_{\rm
  m}\approx 276.0$\,K. While an increase of approximately $3$\,K is
modest, it is nonetheless comparable to the difference in melting
temperature between \ce{D2O} and \ce{H2O}
\cite{bartholome1935calorische,reinhardt2021quantum}.

We have not reported an error estimate for either
$T^{(8.5\to\infty)}_{\rm m}$ or $T_{\rm m}^{(8.5)}$. Yet, the
similarity of the slopes for $\beta\mu_{\rm liq}$ and $\beta\mu_{\rm
  ice}$ seen in Fig.~\ref{fig:BetaMuVsTemp_TailShift} suggest that
even small statistical errors in the chemical potential will result in
relatively large changes in the estimate of the melting
temperature. Instead of performing a thorough error analysis, in
Sec.~\ref{sec:Tm-MF} we use a combination of a MF approach and
Hamiltonian Gibbs-Duhem integration
\cite{agrawal1995solid,agrawal1995thermodynamic} to argue that the
difference in $T_{\rm m}$ reported above reflects a genuine effect of
the choice of truncation schemes.

\section{A mean field estimate for $r_{\rm c}$ dependence of $T_{\rm m}$}
\label{sec:Tm-MF}

We have already seen in Fig.~\ref{fig:RhoVsPress_300K}a that the
density of the homogeneous system under isothermal-isobaric conditions
is sensitive to the choice of $U^{(r_{\rm c})}$ vs. $U^{(r_{\rm
    c}\to\infty)}$. As indicated by the solid blue line,
$\bar{\rho}_{\rm m}(p)$ is well-described by a quadratic polynomial
$r_2p^2 + r_1p + r_0$ \tcr{(see SM)}. Using this polynomial
approximation in combination with Eqs.~\ref{eqn:DelMF-P}
and~\ref{eqn:Pdiff}, we can predict the pressure difference between
the $U^{(r_{\rm c})}$ and $U^{(r_{\rm c}\to\infty)}$ systems, as shown
by the dashed lines in Fig.~\ref{fig:RhoVsPress_300K}a. To validate
this MF estimate, we have performed $NpT$ simulations for
TIP4P/ice$^{(8.5)}$ at $p=p^{(8.5)}$ predicted by
Eq.~\ref{eqn:Pdiff}. Excellent agreement between the simulation data
and MF estimate is observed. This result is perhaps unsurprising, and
simply reflects that $r_{\rm c} = 8.5\,{\rm \AA} \approx 2.7\sigma$ is
sufficiently large to ensure $g_{\rm OO}(r_{\rm c})\approx
1$. Nonetheless, it serves as an acute reminder of the effects of the
truncation scheme: $\bar{\rho}_{\rm m}(p=0)$ for TIP4P/ice$^{(8.5)}$
corresponds to $p\approx -427$\,bar for TIP4P/ice$^{(8.5\to\infty)}$;
even for a relatively large cutoff $r_{\rm c}=14\,{\rm \AA} \approx
4.4\sigma$, differences on the order 100\,bar persist.

\begin{figure}[tb]
  \centering
  \includegraphics[width=8cm]{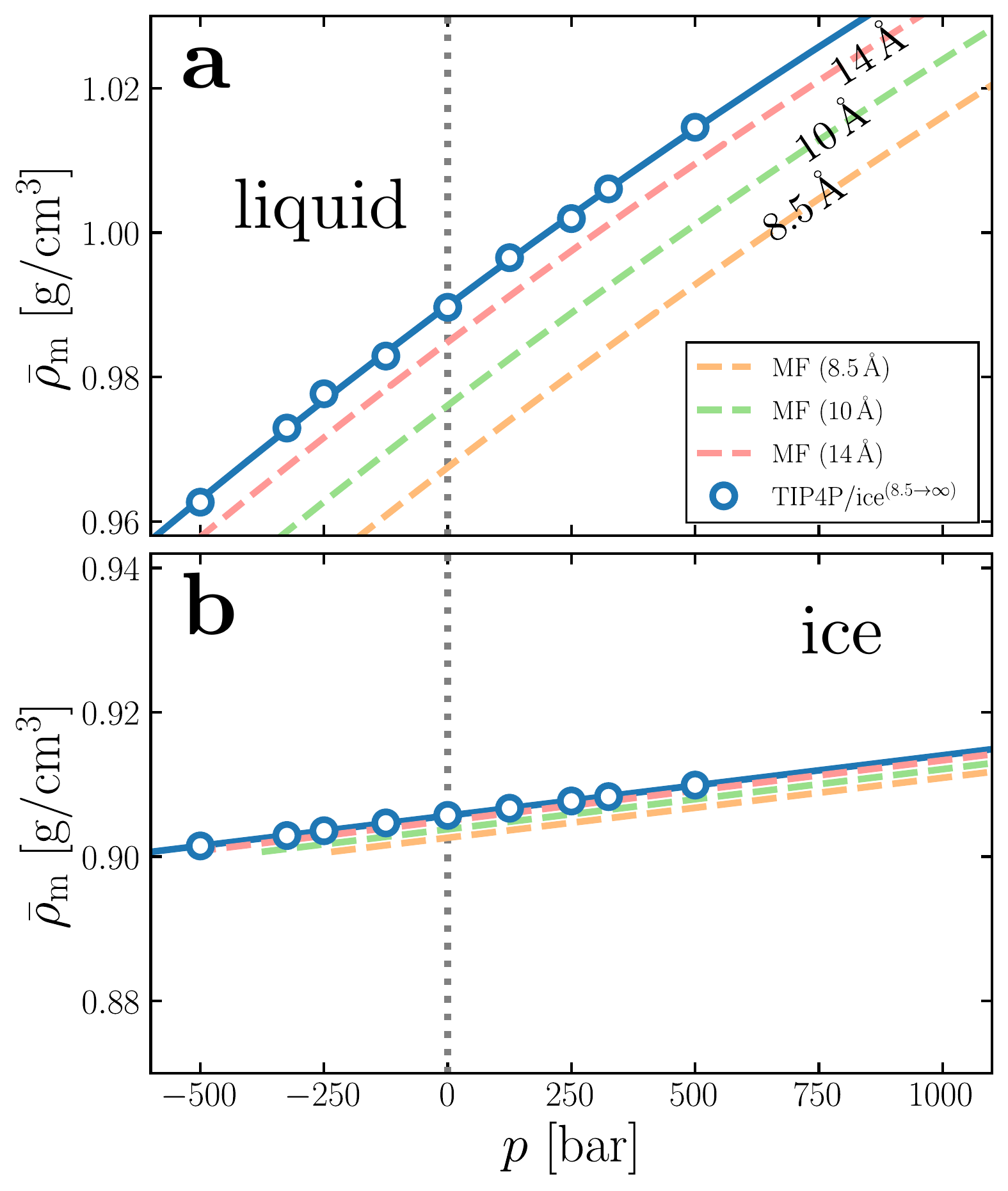}
  \caption{$\bar{\rho}_{\rm m}(p)$ at 272\,K for (a) liquid water and
    (b) ice. White-filled circles show results from constant-$p$
    simulations of TIP4P/ice$^{(8.5\to\infty)}$, and the solid blue
    line indicates a quadratic fit. Dashed lines indicate MF
    predictions (Eq.~\ref{eqn:Pdiff}) for different $r_{\rm c}$, as
    indicated in the legend, which are used to predict
    $\bar{\rho}_{\rm m}(p=0)$ for a given $r_{\rm c}$, i.e., where the
    dashed lines intersect the vertical gray dotted line.}
  \label{fig:RhoVsPress_Tm}
\end{figure}

\begin{figure}[tb]
  \centering
  \includegraphics[width=8cm]{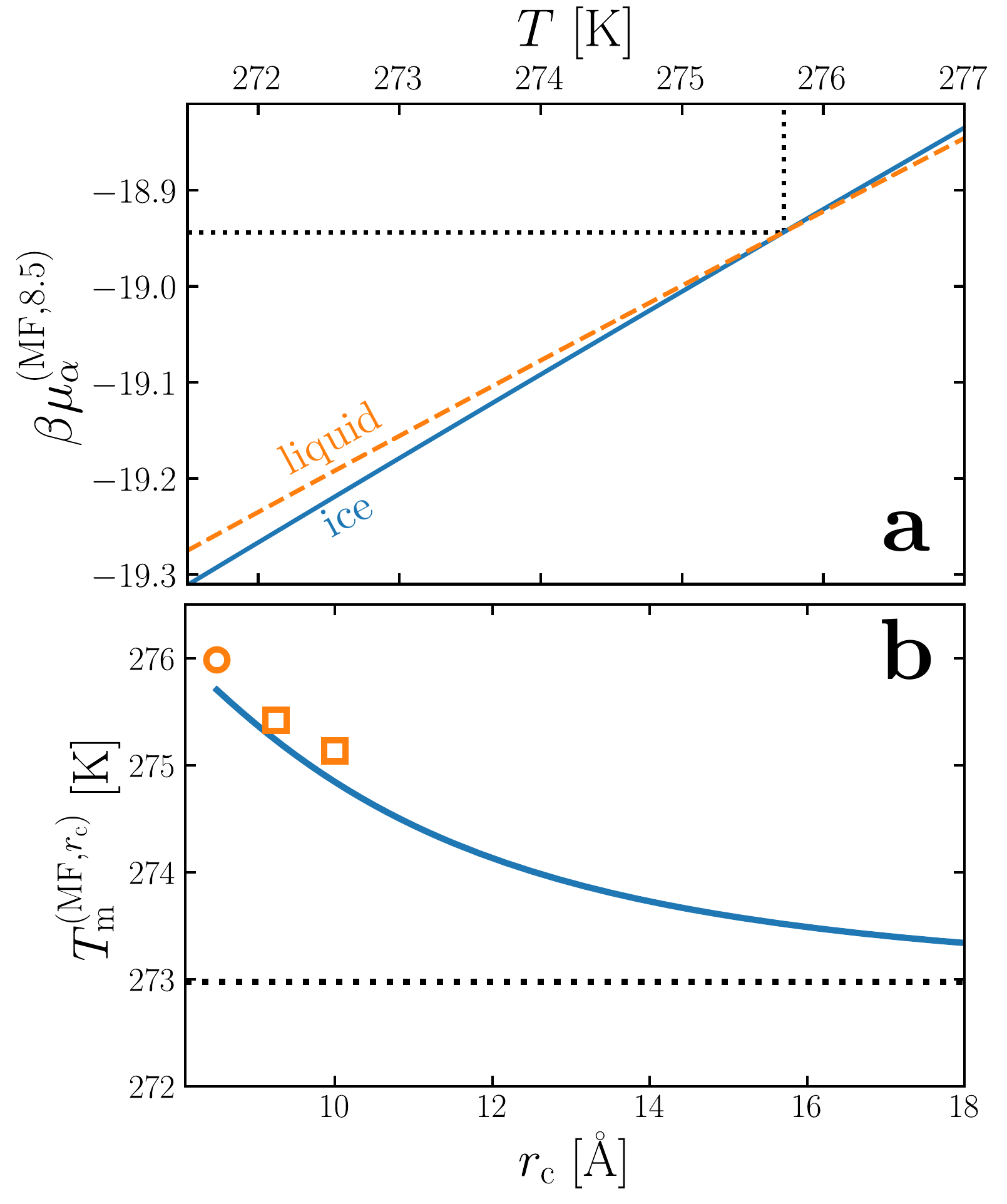}
  \caption{Predicting the effect of $r_{\rm c}$ on the melting
    temperature of TIP4P/ice with MF theory. (a)
    $\beta\mu_\alpha^{({\rm MF,} 8.5)}(T)$ at $p=0$\,bar, with $\alpha
    = \text{`ice' or `liq'}$, obtained from
    Eq.~\ref{eqn:mudiff}. $T_{\rm m}^{({\rm MF,}8.5)}=275.7$\,K is
    determined from the point of interception, as indicated by the
    black dotted lines. (b) $T_{\rm m}^{({\rm MF,}r_{\rm c})}$ is
    shown by the solid blue line. The orange circle indicates $T_{\rm
      m}^{(8.5)}$ obtained from the free energy calculations described
    in Sec.~\ref{sec:meltingEinstein}, and the orange squares indicate
    $T_{\rm m}^{(9.25)}$ and $T_{\rm m}^{(10.0)}$ obtained from
    Hamiltonian Gibbs-Duhem integration, starting from $T_{\rm
      m}^{(8.5)}$.}
  \label{fig:BetaMuVsTemp_MF}
\end{figure}

Assuming that $a_{\rm LJ,ex}^{(r_{\rm c}\to\infty)}\approx a_{\rm
  LJ,ex}^{(\infty)}$, the Helmholtz free energy per particle for a
system with potential energy function $U^{(r_{\rm c})}$ can be
estimated at a MF level\cite{johnson1993lennard},
\begin{equation}
  \label{eqn:aliqrc}
  a_{\rm liq}^{(r_{\rm c})} \approx a_{\rm liq}^{(r_{\rm c}\to\infty)} + \Delta_{\rm MF}a(r_{\rm c}),
\end{equation}
with
\begin{equation}
  \label{eqn:DelMF-a}
  \Delta_{\rm MF}a(r_{\rm c}) =
  -\frac{32\pi\bar{\rho}\epsilon\sigma^3}{9}\Bigg[\bigg(\frac{\sigma}{r_{\rm c}}\bigg)^9-\frac{3}{2}\bigg(\frac{\sigma}{r_{\rm c}}\bigg)^3\Bigg].
\end{equation}
While MF corrections of the kind given by Eqs.~\ref{eqn:DelMF-U},
\ref{eqn:DelMF-P} and~\ref{eqn:DelMF-a} are strictly appropriate for
systems with uniform density, such as homogeneous liquids, they are
often employed for crystalline phases too, with evidence to suggest
that the obtained results are reasonable
\cite{jablonka2019applicability}. (Note that $\Delta_{\rm MF}a$
approximates the difference in free energy between systems employing
$U^{(r_{\rm c}\to\infty)} \approx U^{(\infty)}$ and $U^{(r_{\rm
    c})}$. In contrast, $\Delta_{\rm MF}U$ approximately accounts for
the energy neglected by simply truncating the LJ potential at $r_{\rm
  c}$.) In Figs.~\ref{fig:RhoVsPress_Tm}a and~\ref{fig:RhoVsPress_Tm}b
we show similar analyses as Fig.~\ref{fig:RhoVsPress_300K}a for the
liquid and ice phases of TIP4P/ice, respectively, and temperature
$T_{\rm i}=272$\,K, which allow us to predict $\bar{\rho}_{\rm
  m}(p=0)$ for both phases of TIP4P/ice$^{(r_{\rm c})}$. Along with
Eqs.~\ref{eqn:aliqrc} and~\ref{eqn:DelMF-a}, this estimate of the
density for a given cutoff provides a MF estimate of the chemical
potential:
\begin{equation}
  \label{eqn:mudiff}
  \beta\mu^{({\rm MF},r_{\rm c})} \approx \beta\mu^{(r_{\rm c}\to\infty)}+\Delta_{\rm MF}a(r_{\rm c}).
\end{equation}
Note that, for simplicity, we have ignored any variation of the
density with temperature. Results for $\beta\mu^{({\rm MF},8.5)}$ are
shown in Fig.~\ref{fig:BetaMuVsTemp_MF}, from which we deduce a MF
estimate for the melting temperature $T_{\rm m}^{({\rm MF},8.5)} =
275.7$\,K; this is in fair agreement with $T_{\rm m}^{(8.5)} =
276.0$\,K obtained from our free energy calculations.

Without performing further simulations, we can use the above procedure
to calculate $T^{({\rm MF},r_{\rm c})}_{\rm m}$ for arbitrary $r_{\rm
  c}$, as shown in Fig.~\ref{fig:BetaMuVsTemp_MF}b. It can be clearly
seen that $T^{({\rm MF},r_{\rm c})}_{\rm m}$ approaches $T^{({\rm
    MF},r_{\rm c}\to\infty)}_{\rm m}$ monotonically and slowly, with
differences of approximately 1\,K still observed for $r_{\rm
  c}=12\,{\rm\AA}\approx 3.8\sigma$. Also shown in
Fig.~\ref{fig:BetaMuVsTemp_MF}b are estimates of $T^{(9.25)}_{\rm
  m}=275.4$\,K and $T^{(10.0)}_{\rm m}=275.1$\,K obtained from
Hamiltonian Gibbs-Duhem integration, starting from $T^{(8.5)}_{\rm
  m}=276.0$\,K. The observed relative decrease in $T_{\rm m}$ obtained
from Hamiltonian Gibbs-Duhem integration agrees well with that
predicted by our MF procedure, and provides compelling evidence that
reducing $r_{\rm c}$ results in a systematic increase in the melting
temperature. As already mentioned, the increase in $T_{\rm m}$ with
decreasing $r_{\rm c}$ is modest. We argue that this is a useful
observation, as obtaining consistent ice nucleation rates among
different studies has proven itself to be
challenging\cite{sosso2016crystal}. Our finding suggests that changes
in the degree of supercooling due to differences in $r_{\rm c}$ are an
unlikely source of significant discrepancies in nucleation rates
between studies. In Sec.~\ref{sec:Nucleation}, we suggest a way in
which effects of the truncation scheme can have a material impact on
comparing nucleation rates.

\section{Estimating the impact on ice nucleation rates}
\label{sec:Nucleation}

Our results so far indicate that a finite cutoff results in an
increase, albeit small, on the melting temperature of SPC models of
water. Despite this relatively modest effect on $T_{\rm m}$, we
nonetheless anticipate that the resulting inconsistencies observed
between homogeneous and inhomogeneous systems may have a significant
impact when comparing nucleation rates. In particular,
Figs.~\ref{fig:RhoVsPress_300K} and~\ref{fig:RhoVsPress_Tm} suggest
that a decrease in $r_{\rm c}$ is analogous to an increase in pressure
(for fixed $\bar{\rho}$). Conversely, for inhomogeneous systems like
those shown in Fig.~\ref{fig:Schematic}b, where $U^{(r_{\rm
    c}\to\infty)}$ and $U^{(r_{\rm c})}$ generate the same dynamics,
it is more appropriate to compare to homogeneous nucleation rates
computed with $U^{(r_{\rm c}\to\infty)}$ at $p<0$\,bar rather than
$p=0$\,bar.\footnote{It would, of course, be most appropriate to
  compare to homogeneous nucleation rates obtained with $U^{(r_{\rm
      c})}$ at $p=0$\,bar. Such reference data are, however, scarce.}

\begin{figure}[tb]
  \includegraphics[width=8cm]{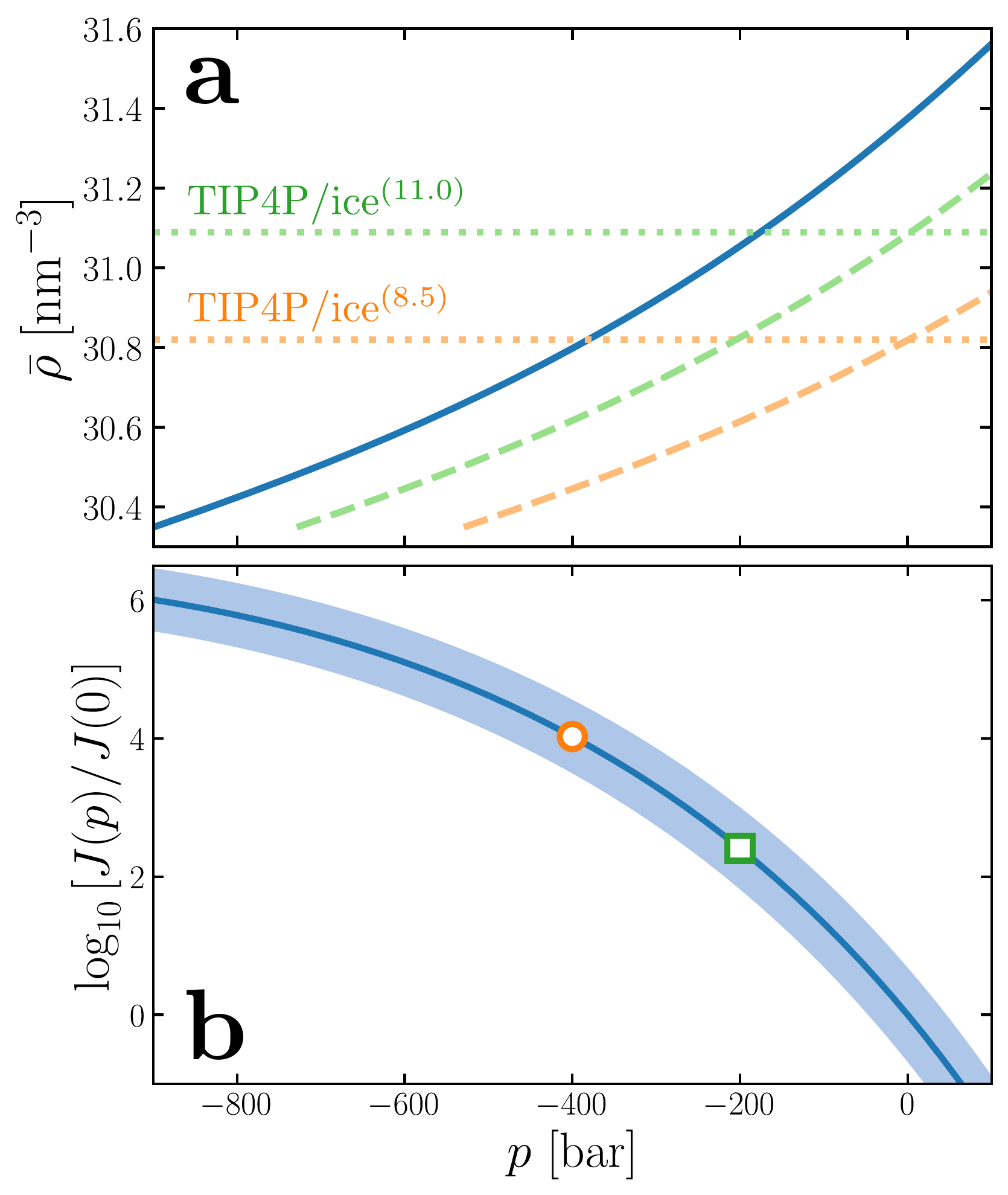}
  \caption{Estimating the impact on ice nucleation rates. (a)
    $\bar{\rho}(p)$ for homogeneous liquid water at 230\,K. The solid
    blue line is the result for
    TIP4P/ice$^{(9.0\to\infty)}$\cite{bianco2021anomalous}. Dashed
    lines indicate MF predictions (Eq.~\ref{eqn:Pdiff}) for
    TIP4P/ice$^{(8.5)}$ (orange) and TIP4P/ice$^{(11.0)}$ (green); the
    dotted lines indicate $\bar{\rho}(p=0)$ for these two
    cut-and-shift variants. For TIP4P/ice$^{(8.5)}$ and
    TIP4P/ice$^{(11.0)}$, $\bar{\rho}(p=0)$ respectively corresponds
    to $p\approx -400$\,bar and $p\approx -200$\,bar for
    TIP4P/ice$^{(9.0\to\infty)}$. (b) $\log_{10}[J(p)/J(0)]$ extracted
    from Ref.~\onlinecite{bianco2021anomalous} with $|\Delta\mu| =
    0.62$\,kJ/mol. At $p\approx -400$\,bar (orange circle) and
    $p\approx -200$\,bar (green square), homogeneous nucleation is
    approximately four and two orders of magnitude faster,
    respectively, than at $p=0$\,bar. }
  \label{fig:Nucleation}
\end{figure}

To estimate the impact of this effective change in pressure arising
from a finite cutoff, we appeal to the recent study of Bianco
\etal\cite{bianco2021anomalous}, where homogeneous nucleation across a
broad range of pressures and temperatures for
TIP4P/ice$^{(9.0\to\infty)}$ was investigated, and data for
$\bar{\rho}(p)$, diffusion coefficient $D(p)$, and size of critical
cluster $n_{\rm c}(p)$ were given. The homogeneous nucleation rate can
then be estimated by
\begin{equation}
  \label{eqn:Jcnt}
  J(p) = \bar{\rho}f^{+}\mathcal{Z}\exp\big(-\beta\Delta G_{\rm c}\big),
\end{equation}
where $\mathcal{Z} = \sqrt{\beta|\Delta\mu|/(6\pi n_{\rm c})}$ and
$f^{+} = 24 Dn_{\rm c}^{2/3}/(3.8\,{\rm \AA})^2$. For simplicity, we
have assumed $|\Delta\mu|=0.62$\,kJ/mol (see Fig.\,3a of
Ref.~\onlinecite{bianco2021anomalous}), independent of pressure; this
is justified based on previous studies that find changes in ice/water
interfacial tension dominate variations in $J$ with $p$, and is
supported by our finding that $T_{\rm m}$ is only weakly affected by
$r_{\rm c}$ \cite{espinosa2016interfacial,bianco2021anomalous}. (To
gauge the sensitivity of our results to this approximation, the blue
shaded region in Fig.~\ref{fig:Nucleation}b encompasses predictions
obtained with $0.60\,{\rm kJ/mol} \le |\Delta\mu| \le 0.64$\,kJ/mol.)
In Fig.~\ref{fig:Nucleation}a we show $\bar{\rho}(p)$ at $T=230$\,K
for TIP4P/ice$^{(r_{\rm c}\to\infty)}$ from
Ref.~\onlinecite{bianco2021anomalous}, along with MF estimates for
TIP4P/ice$^{(8.5)}$ and TIP4P/ice$^{(11.0)}$. From
Fig.~\ref{fig:Nucleation}a, it can clearly be seen that
$\bar{\rho}(p=0)$ for TIP4P/ice$^{(8.5)}$ and TIP4P/ice$^{(11.0)}$
correspond to $p\approx -400$\,bar and $p\approx -200$\,bar,
respectively. In Fig.~\ref{fig:Nucleation}b, we plot
$\log_{10}\big[J(p)/J(0)\big]$ according to Eq.~\ref{eqn:Jcnt}, from
which we estimate that homogeneous nucleation is faster in
TIP4P/ice$^{(8.5)}$ and TIP4P/ice$^{(11.0)}$ by approximately four and
two orders of magnitude, respectively.

The implication of the preceding analysis is that enhancement due to
heterogeneous nucleation may in fact be lower than previously
thought. For example, Sosso \etal\cite{sosso2016ice} used a variation
of the cut-and-shift potential\footnote{To be concrete, Sosso
  \etal\cite{sosso2016ice} considered LJ interactions up to 10\,\AA,
  where a switching function was used to bring them to zero at
  12\,\AA. Without validation, we simply approximate this by
  TIP4P/ice$^{(11.0)}$.} with $r_{\rm c}\approx 11$\,\AA{} to
investigate ice nucleation at 230\,K in the presence of kaolinite,
using FFS and TIP4P/ice. By comparing to the homogeneous nucleation
rate obtained by Haji-Akbari and Debenedetti for
TIP4P/ice$^{(8.5\to\infty)}$ with FFS, an enhancement of 20 orders of
magnitude was reported; we estimate this result is too high by
approximately two orders of magnitude. Similarly, Haji-Akbari and
Debenedetti also investigated nucleation in free standing thin films
of TIP4P/ice$^{(8.5)}$ water \cite{haji2017computational} and found an
increase of approximately seven orders of magnitude, despite
nucleation occurring in bulk-like regions; Fig.~\ref{fig:Nucleation}b
suggests the nucleation rate of the reference homogeneous system at $p
= -400$\,bar would also be faster by approximately four orders of
magnitude.

This discussion on the impact of truncation scheme on the nucleation
rate is admittedly crude, and relies on the analogy that a change in
$r_{\rm c}$ simply amounts to a change in pressure. In practice, it is
likely that relevant quantities, e.g., ice-liquid interfacial tension,
will differ between TIP4P/ice$^{(r_{\rm c}\to\infty)}$ at $p<0$\,bar
and TIP4P/ice$^{(r_{\rm c})}$ at $p=0$\,bar. While the estimates
presented above may provide a useful first-order approximation, they
await full validation by explicit calculation of nucleation rates
using consistent truncation schemes for homogeneous and inhomogeneous
systems. Such calculations are, however, beyond the scope of the
present article.

\section{Summary and outlook}
\label{sec:concl}

In this article, we have investigated the effect of truncating the
Lennard-Jones potential on the melting properties at $p=0$\,bar of two
common water models---TIP4P/ice and TIP4P/2005---that are frequently
used to study ice nucleation with molecular simulations. Specifically,
we have compared results from two truncation schemes: simple
truncation at $r_{\rm c}$ with `tail corrections'; and `cut-and-shift'
at $r_{\rm c}$. We have combined explicit free energy calculations,
Hamiltonian Gibbs-Duhem integration, and a simple mean field analysis
to show that a finite cutoff results in an increase of the melting
temperature. While we have focused on TIP4P/ice and TIP4P/2005, the
effects described in this article should be applicable to any
reasonable SPC model of water. Moreover, while not an SPC model, we
note that the coarse grained mW
model\cite{molinero2009water}---another water model commonly used to
investigate ice nucleation---is inherently short-ranged, with
intermolecular interactions that vanish beyond 4.32\,\AA. As such, we
can conclude that the mW model will not suffer from the
inconsistencies between homogeneous and inhomogeneous systems
discussed in this article.

Based on recent work that has investigated homogeneous ice nucleation
at negative pressures \cite{bianco2021anomalous}, we suggest that
enhancements due to heterogeneous nucleation calculated by molecular
simulations have likely been overestimated by several orders of
magnitude. Going forward, those simulating heterogeneous nucleation
either need to employ a truncation scheme that effectively samples
$U^{(r_{\rm
    c}\to\infty)}$\cite{in2007application,alejandre2010surface,lopez2002thermodynamic,lopez2003simulation,miguez2013influence,janecek2006long,salomons1991atomistic,guo1997long1,guo1997long2,de2012semi},
or reference data for homogeneous nucleation rates for $U^{(r_{\rm
    c})}$-based SPC models needs to be computed explicitly. As a
stop-gap solution, one can use the crude but cheap estimate for the
impact on comparing homogeneous and heterogeneous nucleation rates
outlined in this article.

Inconsistencies arising from the choice of truncation scheme are not
the only challenges faced when comparing homogeneous and heterogeneous
ice nucleation. In particular, we note that Haji-Akbari has shown that
conventional FFS approaches can underestimate nucleation rates by
failing to account for the `jumpiness' of the order parameter, the
severity of which is system dependent \cite{haji2018forward}. While
such subtleties in rate calculations further complicate quantitative
comparison of homogeneous and heterogeneous nucleation rates, our work
nonetheless provides an important contribution toward resolving
inconsistencies between homogeneous and inhomogeneous systems. Our
results will also facilitate consistent comparison of different
studies of heterogeneous ice nucleation.

\section{Methods}
\label{sec:methods}

Full details of the methods used are given in the SM. In brief,
molecular dynamics simulations were performed with the \texttt{LAMMPS}
simulations package \cite{plimpton1995sjc}. The particle-particle
particle-mesh Ewald method was used to account for long-ranged
interactions \cite{HockneyEastwood1988sjc}, with parameters chosen
such that the root mean square error in the forces were a factor
$10^{5}$ smaller than the force between two unit charges separated by
a distance of 0.1\,nm \cite{kolafa1992cutoff}. For simulations of a
liquid water slab in contact with its vapor, the electric displacement
field along $z$ was set to zero, using the implementation given in
Refs.~\onlinecite{cox2019finite,sayer2019stabilization}; this is
formally equivalent to the commonly used slab correction of Yeh and
Berkowitz \cite{YehBerkowitz1999sjc}. The geometry of the water
molecules was constrained using the \texttt{RATTLE} algorithm
\cite{andersen1983rattle}. Where appropriate, temperature was
maintained with either a Nos\'{e}-Hoover chain thermostat
\cite{shinoda2004rapid,tuckerman2006liouville} or Langevin dynamics
\cite{DunwegWolfgang1991sjc,SchneiderStoll1978sjc}, and pressure with
a Parrinello-Rahman barostat \cite{parrinello1981polymorphic} with a
damping constant 2\,ps. A time step of 2\,fs was used throughout. Ice
structures were generated using the GenIce software package
\cite{matsumoto2018genice}.

\section*{Supplementary Material}
Supplementary Material includes a detailed overview of the simulation
methods used. Results for the TIP4P/2005 water model are also given.

\begin{acknowledgments}
We are grateful to Aleks Reinhardt, Christoph Schran, Martin Fitzner
and Gabriele Sosso for comments on our manuscript. Amir Haji-Akbari
and Pablo Debenedetti are thanked for providing information on their
simulations. \tcb{We thank Daan Frenkel for insightful discussions
  concerning impulsive forces.} S.J.C is a Royal Society University
Research Fellow (URF\textbackslash R1\textbackslash 211144) at the
University of Cambridge.
\end{acknowledgments}

\section*{Data Availability Statement}

The data that supports the findings of this study, analysis scripts,
and input files for the simulations are openly available at the
University of Cambridge Data Repository,
\url{https://doi.org/10.17863/CAM.80092}.

\bibliography{../urf}

\clearpage
\onecolumngrid
\renewcommand\thefigure{S\arabic{figure}}
\renewcommand\theequation{S\arabic{equation}}
\renewcommand\thesection{S\arabic{section}}
\setcounter{figure}{0}
\setcounter{equation}{0}
\setcounter{section}{0}

\noindent {\Large \tbf{Supplementary Material}}

\section{Background theory for calculating the free energy of the liquid and crystalline phases}

\noindent To help set notation, and highlight slight differences in
approach compared to previous studies, we will briefly cover some of
the theory underlying the free energy calculations performed in the
main article.

\subsection{Liquid}

\noindent As we consider rigid water molecules, the position of all
atoms in molecule $i$ can be specified entirely by the location of its
oxygen atom $\mbf{r}_i^{\rm (O)}\equiv \mbf{R}_i$ and its orientation
$\bm{\Omega}_i$. The translational and rotational momentum of molecule
$i$ are denoted $\mbf{p}_i$ and $\mbf{L}_i$, respectively. The
partition function for a system comprising $N$ indistinguishable
molecules can thus be written as
\begin{align}
  Q &= \frac{1}{h^{6N}N!}\int\!\mrm{d}\mbf{p}^{N}\,\int\!\mrm{d}\mbf{L}^{N}\,
  \int\!\mrm{d}\mbf{R}^{N}\int\!\mrm{d}\mbf{\Omega}^{N}
  \me^{-\beta\mathcal{K}_{\rm t}(\mbf{p}^N)}\me^{-\beta\mathcal{K}_{\rm r}(\mbf{L}^N)}\me^{-\beta\mathcal{U}(\mbf{R}^N,\bm{\Omega}^N)}, \\[7pt]
  &= \frac{(8\pi^2)^NV^N}{h^{6N}N!}\int\!\mrm{d}\mbf{p}^{N}\,\me^{-\beta\mathcal{K}_{\rm t}(\mbf{p}^N)}
  \int\!\mrm{d}\mbf{L}^{N}\,\me^{-\beta\mathcal{K}_{\rm r}(\mbf{L}^N)}
  \frac{1}{V^N}\int\!\mrm{d}\mbf{R}^{N}\frac{1}{(8\pi^2)^N}\int\!\mrm{d}\mbf{\Omega}^{N}
  \me^{-\beta\mathcal{U}(\mbf{R}^N,\bm{\Omega}^N)},
\end{align}
where $h^{6N}$ defines a volume element in phase space,
$\mathcal{K}_{\rm t}$ and $\mathcal{K}_{\rm r}$ are the translational
and rotational kinetic energy, respectively, and $\mathcal{U}$ is the
potential energy. For non-linear rigid molecules like the water models
considered, 
\begin{equation}
  \mathcal{K}_{\rm r}(\mbf{L}) =
  \sum_{i=1}^{N}\frac{|\mbf{L}_{i}^{(1)}|^2}{2I^{(1)}}+\frac{|\mbf{L}_{i}^{(2)}|^2}{2I^{(2)}}+\frac{|\mbf{L}_{i}^{(3)}|^2}{2I^{(3)}},
\end{equation}
where the superscripts indicate different principal axes of rotation,
and $I^{(1)}$ indicates the moment of inertia around axis 1 etc. The
ideal contribution to the partition function is then
\begin{equation}
  Q_{\rm id} = \frac{(8\pi^2)^NV^N}{\Lambda^{3N}N!}
  \bigg(\frac{2\pi k_{\rm B}TI^{(1)}}{h^2}\bigg)^{N/2}\bigg(\frac{2\pi k_{\rm B}TI^{(2)}}{h^2}\bigg)^{N/2}\bigg(\frac{2\pi k_{\rm B}TI^{(3)}}{h^2}\bigg)^{N/2}.
\end{equation}
If the total mass of a molecule is $m$, then we can write e.g.,
\[
\bigg(\frac{2\pi mk_{\rm B}T}{h^2}\bigg)^{N/2}\bigg(\frac{I^{(1)}}{m}\bigg)^{N/2} = \frac{1}{\Lambda^N}\bigg(\frac{I^{(1)}}{m}\bigg)^{N/2}.
\]
Thus,
\begin{align}
  \ln Q_{\rm id} \approx &N\ln\bigg(\frac{V}{N\Lambda^3}\bigg) + N \nonumber \\[7pt]
  +&N\ln\bigg[\bigg(\frac{I^{(1)}}{m}\bigg)^{1/2}\frac{1}{\Lambda}\bigg]
  + N\ln\bigg[\bigg(\frac{I^{(2)}}{m}\bigg)^{1/2}\frac{1}{\Lambda}\bigg]
  + N\ln\bigg[\bigg(\frac{I^{(3)}}{m}\bigg)^{1/2}\frac{1}{\Lambda}\bigg] + N\ln 8\pi^2 \\[7pt]
  &= -N\ln\big(\bar{\rho}\Lambda^3\big) + N -N\ln\bigg[\bigg(\frac{m^3}{I^{(1)}I^{(2)}I^{(3)}}\bigg)^{1/2}\frac{\Lambda^3}{8\pi^2}\bigg], \\[7pt]
  &= -N\ln\big(\bar{\rho}\eta_{\rm r}\Lambda^6\big) + N,
\end{align}
where we have defined
\[
\eta_{\rm r} \equiv \bigg(\frac{m^3}{I^{(1)}I^{(2)}I^{(3)}}\bigg)^{1/2}\frac{1}{8\pi^2}.
\]

Let us now write $\eta_{\rm r}^{1/6}\Lambda =
\bar{\Lambda}_0T^{-1/2}$, such that
$\bar{\Lambda}_0T_0^{-1/2}=1$\,\AA$^{1/2}$. Then,
\[
\eta_{\rm r}^{1/6}\Lambda = \bar{\Lambda}_0T_0^{-1/2}\bigg(\frac{T_0}{T}\bigg)^{1/2} = \bigg(\frac{T_0}{T}\bigg)^{1/2}\,\text{\AA}^{1/2}.
\]
The ideal free energy can then be written as,
\begin{equation}
  \frac{\beta A_{\rm id}}{N} = \beta a_{\rm id} = \ln\big(\bar{\rho}(T_0/T)^3\big) - 1,
\end{equation}
where it is understood that $(T_0/T)$ carries units of \AA. The choice
of reference temperature $T_0$ is arbitrary provided it is chosen
consistently. This approach differs from the common `set
$\Lambda=1$\,\AA' encountered in the literature
\cite{vega2008determination}. By adopting this approach we use, e.g.,
the full enthalpy when performing thermodynamic integration
(c.f. Ref.~\onlinecite{reinhardt2019phase}).

The excess part of the partition function is
\begin{equation}
Q_{\rm ex} = \frac{1}{(8\pi^2V)^N}\int\!\mrm{d}\mbf{R}^N\int\!\mrm{d}\mbf{\Omega}^{N}
  \me^{-\beta\mathcal{U}(\mbf{R}^N,\bm{\Omega}^N)}.
\end{equation}
Note that, if $\mathcal{U}$ is independent of $\Omega$ e.g., we turn
off the charges in our water model, then $Q_{\rm ex}$ reduces to that
of a simple mono-atomic system. This means we are free to use
equations of state for the standard LJ liquid where appropriate; we
make use of this fact to calculate the excess free energy of the
liquid by thermodynamic integration.

\subsection{Ice}
\label{subsec:Theory:Ice}

Unlike liquid water, the molecules in the crystalline phase are
distinguishable by virtue of their association with a particular set
of lattice sites. This leads to a straightforward modification of the
partition function:
\begin{equation}
    Q = \frac{1}{h^{6N}}\int\!\mrm{d}\mbf{p}^{N}\,\int\!\mrm{d}\mbf{L}^{N}\,
  \int\!\mrm{d}\mbf{R}^N\int\!\mrm{d}\mbf{\Omega}^{N}
  \me^{-\beta\mathcal{K}_{\rm t}(\mbf{p}^N)}\me^{-\beta\mathcal{K}_{\rm r}(\mbf{L}^N)}\me^{-\beta\mathcal{U}(\mbf{R}^N,\bm{\Omega}^N)}
\end{equation}
Instead of dealing with `ideal' and `excess' quantities, it is now
useful to consider `kinetic' and `configurational' quantities:
\begin{align}
  Q_{\rm kin} &= \frac{1}{\Lambda^{6N}\eta_{\rm r}^N}, \\[7pt]
  Q_{\rm con} &= \frac{1}{(8\pi^2)^N}\int\!\mrm{d}\mbf{r}_{\rm O}^{N}\int\!\mrm{d}\mbf{\Omega}^{N}
  \me^{-\beta\mathcal{U}(\mbf{r}_{\rm O}^N,\bm{\Omega}^N)}.
\end{align}
Note that $Q_{\rm kin}$ and $Q_{\rm con}$ have dimensions of
hyperdensity and hypervolume, respectively; it is important that units
are chosen consistently. The factor $1/(8\pi^2)^N$ is still included
in $Q_{\rm con}$ to ensure a consistent definition of $\eta_r$. By
similar reasoning to above, we can write the kinetic contribution to
the free energy as
\begin{equation}
  \frac{\beta A_{\rm kin}}{N} = \beta a_{\rm kin} = \ln\big((T_0/T)^3\big).
\end{equation}

As detailed below, we have used the Frenkel-Ladd approach
\cite{frenkel1984new}, adapted by Vega and co-workers for rigid SPC
water models
\cite{noya2008computing,vega2008determination,aragones2013free}, to
calculate the difference in free energy between a non-interacting
crystal with its atoms tethered to their equilibrium positions by
harmonic springs, and the fully interacting crystal. The potential
energy of the former, `reference', system is
\begin{equation}
  \label{eqn:Uref}
  \mathcal{U}_{\rm ref}(\mbf{R}^N,{\bm\Omega}^N) =
  \sum_i^N\sum_{\alpha} \frac{k^{(\alpha)}}{2}\big(\mbf{R}_i+\Delta\mbf{r}_i^{(\alpha)}({\bm\Omega_i})-\mbf{r}_i^{(\alpha,0)}\big)^2,
\end{equation}
where $\Delta\mbf{r}_i^{(\alpha)} = \mbf{r}_i^{(\alpha)}-\mbf{R}_i$,
$\mbf{r}_i^{(\alpha,0)}$ is the equilibrium position of atom $\alpha$
of molecule $i$ (recall that $\mbf{r}_i^{\rm (O)}\equiv \mbf{R}_i$),
and $k^{(\alpha)}$ determines the strength of the harmonic potential
that tethers atom $\alpha$ to $\mbf{r}_i^{(\alpha,0)}$. The rigid body
constraints mean that the free energy of this reference system is
analytically intractable. We therefore define a `sub-reference' system
with the following potential energy,
\begin{equation}
  \label{eqn:Usub}  
  \mathcal{U}_{\rm sub}(\mbf{R}^N) =
  \sum_i^N\frac{k^{\rm (O)}}{2}\big(\mbf{r}^{\rm (O)}_i-\mbf{r}_i^{({\rm O},0)}\big)^2.
\end{equation}
The configurational partition function for this sub-reference system
is just that of the standard Einstein crystal,
\begin{equation}
  Q_{\rm sub} = \int\!\mrm{d}\mbf{R}^N \exp\left(-\beta \mathcal{U}_{\rm sub}(\mbf{R}^N)\right), 
\end{equation}
resulting in the following free energy per particle:
\begin{equation}
  \beta a_{\rm sub} = -\frac{3}{2}\ln\bigg(\frac{2\pi}{\beta k^{\rm (O)}}\bigg).
\end{equation}

\section{Workflow: Free energy calculations of ice I$_{\rm h}$}
\label{sec:WorkflowIce}

\noindent The procedure described below was performed for both
truncation schemes described in the main article and both water
models, i.e., for TIP4P/ice$^{(8.5\to\infty)}$, TIP4P/ice$^{(8.5)}$,
TIP4P/2005$^{(8.5\to\infty)}$, and TIP4P/2005$^{(8.5)}$. Unless
otherwise stated, all simulations used the \texttt{LAMMPS} simulation
package \cite{plimpton1995sjc}. The particle-particle particle-mesh
Ewald method was used to account for long-ranged interactions
\cite{HockneyEastwood1988sjc}, with parameters chosen such that the
root mean square error in the forces were a factor $10^{5}$ smaller
than the force between two unit charges separated by a distance of
0.1\,nm \cite{kolafa1992cutoff}. The geometry of the water molecules
was constrained using the \texttt{RATTLE} algorithm
\cite{andersen1983rattle}. A time step of 2\,fs was used throughout.

\subsection{Obtaining average cell parameters}
\label{subsec:WorkflowIce:AvgCell}

\noindent A proton disordered ice I$_{\rm h}$ structure comprising 768
molecules was generated using the GenIce software package
\cite{matsumoto2018genice}. After equilibration of at least 0.5\,ns,
the average cell parameters were obtained from a 10\,ns simulation at
$p=0$\,bar and temperature $T=T_{\rm i}$, with $T_{\rm i}=272$\,K for
TIP4P/ice, and $T_{\rm i}=252$\,K for TIP4P/2005. Temperature was
maintained with a Nos\'{e}-Hoover chain thermostat
\cite{shinoda2004rapid,tuckerman2006liouville} with a damping constant
0.2\,ps, and the pressure was maintained with a Parrinello-Rahman
barostat \cite{parrinello1981polymorphic} with a damping constant
2\,ps. The latter was applied such that all cell lengths and angles
could fluctuate independently.

\subsection{Obtaining the reference ice structure}

\noindent The simulation cell parameters were fixed to their average
values, and the structure was `minimized' by running short
(approximately 10-20\,ps) simulations at $T=0.1$\,K. The damping
constant of the Nos\'{e}-Hoover chain thermostat was reduced to
20\,fs. As explained in the main text, this approach was adopted as
standard minimizers available in \texttt{LAMMPS} are incompatible with
the \texttt{RATTLE} algorithm used to constrain the rigid geometry of
the water molecules. Simulation settings were otherwise the same as
above.

\subsection{Thermodynamic integration from the non-interacting to interacting crystal}

\noindent Atoms were tethered to their positions in the reference ice
structure with force constants $k^{\rm (O)}=4.8$\,kcal/mol-\AA$^2$ and
$k^{\rm (H)}=6.0$\,kcal/mol-\AA$^2$ (see
Sec.~\ref{subsec:Theory:Ice}). For each water model and truncation
scheme considered, we constructed the following potential energy
function:
\begin{equation}
  \mathcal{U}_\lambda(\mbf{R}^N,{\bm\Omega}^N) = \lambda\mathcal{U}(\mbf{R}^N,{\bm\Omega}^N) + (1-\lambda)\mathcal{U}_{\rm ref}(\mbf{R}^N,{\bm\Omega}^N),
\end{equation}
where $\mathcal{U}$ is replaced with $U^{(r_{\rm c}\to\infty)}$ or
$U^{(r_{\rm c})}$ as appropriate (see Eqs.~\ref{eqn:Utot-MF}
and~\ref{eqn:Utot-cs}). The Helmholtz free energy difference between
the reference and interacting systems is then,
\begin{equation}
  \label{eqn:TI-r2i}
  \Delta_{\rm r2i}a = \frac{1}{N}\int_0^1\!\mrm{d}\lambda\,\langle\Delta U(\mbf{R}^N,{\bm\Omega}^N)\rangle_\lambda,
\end{equation}
where $\Delta U(\mbf{R}^N,{\bm\Omega}^N) =
\mathcal{U}(\mbf{R}^N,{\bm\Omega}^N) - \mathcal{U}_{\rm
  ref}(\mbf{R}^N,{\bm\Omega}^N)$, and $\langle\cdots\rangle_\lambda$
denotes a canonical ensemble average according to the Hamiltonian
specified by $\mathcal{U}_\lambda$. The integral in
Eq.~\ref{eqn:TI-r2i} was evaluated using 11-point Gauss-Legendre
quadrature, and simulations for each value of $\lambda$ were 20\,ns in
length. Temperature was maintained through Langevin dynamics as
implemented in \texttt{LAMMPS}
\cite{DunwegWolfgang1991sjc,SchneiderStoll1978sjc}, with a damping
constant 100\,fs. The total random force was set exactly to zero to
ensure the center-of-mass of the system did not drift.

\subsection{Thermodynamic integration from the sub-reference to reference system}

\noindent As molecules in both the sub-reference and reference systems
are non-interacting, we need only consider the behavior of a single
water molecule. Specifically, we construct the following energy function:
\begin{equation}
  u_\lambda(\mbf{R}_1,{\bm\Omega}_1) = \lambda\mathcal{U}^{(N=1)}_{\rm ref}(\mbf{R}_1,{\bm\Omega}_1) + (1-\lambda)\mathcal{U}^{(N=1)}_{\rm sub}(\mbf{R}_1,{\bm\Omega}_1),
\end{equation}
where $\mathcal{U}^{(N=1)}_{\rm ref}$ and $\mathcal{U}^{(N=1)}_{\rm
  sub}$ are given by Eqs.~\ref{eqn:Uref} and~\ref{eqn:Usub} with
$N=1$. The change in Helmholtz free energy is then given by:
\begin{equation}
  \label{eqn:TI-s2r}
  \Delta_{\rm s2r}a = \int_0^1\!\mrm{d}\lambda\,\langle\Delta u(\mbf{R}^N,{\bm\Omega}^N)\rangle_\lambda,
\end{equation}
with $\Delta u = \mathcal{U}^{(N=1)}_{\rm ref} -
\mathcal{U}^{(N=1)}_{\rm sub}$, and $\langle\cdots\rangle_\lambda$ now
denotes a canonical ensemble average at temperature $T_{\rm i}$
according to the Hamiltonian specified by $u_\lambda$. The integral in
Eq.~\ref{eqn:TI-s2r} was again evaluated using 11-point Gauss-Legendre
quadrature, using a bespoke Metropolis Monte Carlo (MC) code. In
brief, after $10^4$ MC moves for equilibration, production simulations
of $5\times 10^7$ MC moves were performed for each value of
$\lambda$. For each MC move, the water molecule was either translated
or rotated with equal probability. For translations, a displacement
along each Cartesian direction was randomly chosen in the interval
$[-\sqrt{2/\beta k^{\rm (O)}},\sqrt{2/\beta k^{\rm (O)}})$. For
  rotations, three angles ($\alpha,\beta,\gamma$) were randomly chosen
  in the interval $[0,\pi/6)$, and a rotation matrix was constructed
    as $\mbf{R} = \mbf{R}_z(\alpha)\mbf{R}_y(\beta)\mbf{R}_x(\gamma)$,
    where $\mbf{R}_x(\gamma)$ is a rotation about the $x$-axis
    etc. With equal probability, the molecule was then rotated about
    its oxygen position using either $\mbf{R}$ or its transpose. Note
    that, as $\Delta_{\rm s2r}a$ is independent of truncation scheme,
    we only computed it once for each water model.

\subsection{Computing $\beta\mu_{\rm ice}(T)$}
\label{subsec:WorkflowIce:GibbsHelmholtz}

\noindent With an estimate of $\beta_{\rm i}\mu_{\rm ice}(T_{\rm i})$
obtained from thermodynamic integration, $\beta\mu_{\rm ice}(T)$ is
computed from the Gibbs-Helmholtz relation
(Eq.~\ref{eqn:GibbsHelmholtz}). For TIP4P/ice$^{(8.5\to\infty)}$ and
TIP4P/ice$^{(8.5)}$, simulations in the temperature range $T =
267\,{\rm K}, 268\,{\rm K},\ldots,277\,{\rm K}$, and $T = 267\,{\rm
  K}, 268\,{\rm K},\ldots,282\,{\rm K}$, respectively, were performed,
while for TIP4P/2005$^{(8.5\to\infty)}$ and TIP4P/2005$^{(8.5)}$ we
adopted the temperature range $T = 247\,{\rm K}, 248\,{\rm
  K},\ldots,267\,{\rm K}$. Simulations were initialized from the
reference structure, starting at 0.1\,K with the temperature steadily
increased to $T$ over 1\,ns at constant volume. An equilibration
period of 0.5\,ns at constant $T$ and $p=0$\,bar was then performed
(see Sec.~\ref{subsec:WorkflowIce:AvgCell}), followed by a production
run of 20\,ns. The integrand in Eq.~\ref{eqn:GibbsHelmholtz} was then
fitted to a quadratic polynomial, from which $\beta\mu_{\rm ice}(T)$
was obtained by analytic integration.

\section{Workflow: Free energy calculations of liquid water}
\label{sec:WorkflowLiq}

\noindent The procedure described below is again appropriate for both
truncation schemes and both water models. Simulation details were
broadly similar to those specified throughout
Sec.~\ref{sec:WorkflowIce}.

\subsection{Obtaining the average density of liquid water}
\label{subsec:WorkflowLiq:dens}

\noindent A 20\,ns simulation of liquid water was performed after at
least 0.5\,ns equilibration at temperature $T_{\rm i}$ and
$p=0$\,bar. A Nos\'{e}-Hoover chain thermostat was used to maintain
the temperature, and an isotropic Parrinello-Rahman barostat was used
to maintain the pressure.

\subsection{Thermodynamic integration from the LJ fluid to water}

\noindent To compute the excess free energy of liquid water, we
exploit the fact that the equation of state for the LJ fluid has been
computed previously, which provides $a^{(r_{\rm c}\to\infty)}_{\rm
  LJ,ex}$. The density of the fluid is fixed to its average (see
Sec.~\ref{subsec:WorkflowLiq:dens}) at temperature $T_{\rm i}$ and
$p=0$\,bar, and thermodynamic integration is performed with the
following energy function:
\begin{equation}
  \label{eqn:Ulam-liq}
  \mathcal{U}_\lambda(\mbf{R}^N,{\bm\Omega}^N) =
  \mathcal{U}(\mbf{R}^N,{\bm\Omega}^N)\text{ with charges multiplied by $\lambda^{1/2}$.}
\end{equation}
(We reuse the notation $\mathcal{U}_\lambda$ as it should be clear
from context what is intended.) Again, $\mathcal{U}$ is replaced with
$U^{(r_{\rm c}\to\infty)}$ or $U^{(r_{\rm c})}$ as appropriate. The
free energy difference $\Delta_{\rm LJ2w}a$ between water and the LJ
fluid is then given by an expression analogous to
Eq.~\ref{eqn:TI-r2i}, with the integral evaluated by 9-point
Gauss-Legendre quadrature. For each value of $\lambda$, $\langle\Delta
U(\mbf{R}^N,{\bm\Omega}^N)\rangle_\lambda$ was averaged over a 20\,ns
simulation, following a 0.5\,ns equilibration period.

\subsection{Thermodynamic integration from the `truncated + tail corrections' LJ fluid to `cut-and-shift' LJ fluid}

\noindent For systems employing the `cut-and-shift' truncation scheme,
we also computed the free energy difference between the fluid with
interactions described by $u_{\rm LJ}^{(r_{\rm c}\to\infty)}$ and
$u_{\rm LJ}^{(r_{\rm c})}$. As dynamics in the canonical ensemble are
unaffected by this choice of truncation scheme, we simply have (see
Eq.~\ref{eqn:Ulam-liq})
\begin{equation}
  \Delta_{\rm tc2cs}a =
  \big\langle U_{\lambda = 0}^{(8.5)}(\mbf{R}^N,{\bm\Omega}^N)-U_{\lambda = 0}^{(8.5\to\infty)}(\mbf{R}^N,{\bm\Omega}^N)\big\rangle,
\end{equation}
which we calculated from a 20\,ns simulation, following a 0.5\,ns
equilibration period.

\subsection{Computing $\beta\mu_{\rm liq}(T)$}

\noindent Using the same temperature ranges described in
Sec.~\ref{subsec:WorkflowIce:GibbsHelmholtz}, the Gibbs-Helmholtz
equation was evaluated in an analogous manner to $\beta\mu_{\rm
  ice}(T)$. For each temperature, a 0.5\,ns equilibration period was
performed followed by a 20\,ns production run. The pressure was
maintained with an isotropic barostat (see
Sec.~\ref{subsec:WorkflowLiq:dens}).

\section{Workflow: Locating the melting point}

\noindent For each water model and truncation scheme, $\beta\mu_{\rm
  ice}(T)$ and $\beta\mu_{\rm liq}(T)$ were each fitted to a quadratic
polynomial, and the melting temperature was obtained by solving the
resulting simultaneous equations.

\section{Workflow: Hamiltonian Gibbs-Duhem integration}

\noindent With $T_{\rm m}^{(8.5)}$ determined from the free energy
approach described above, $T_{\rm m}^{(9.25)}$ and $T_{\rm
  m}^{(10.0)}$ were subsequently determined by Hamiltonian Gibbs-Duhem
integration. Specifically, we define the potential energy function
\begin{equation}
  U_\lambda(\mbf{R}^N,{\bm\Omega}^N) = \lambda U^{(r_{{\rm
        c},1})}(\mbf{R}^N,{\bm\Omega}^N) + (1-\lambda)U^{(r_{{\rm c},0})}(\mbf{R}^N,{\bm\Omega}^N),
\end{equation}
and the quantity,
\begin{equation}
  x^{(\lambda)}_{\alpha} = \frac{1}{N}\big\langle U^{(r_{{\rm c},1})}(\mbf{R}^N,{\bm\Omega}^N) - U^{(r_{{\rm c},0})}(\mbf{R}^N,{\bm\Omega}^N)\big\rangle_\lambda,
\end{equation}
where $\alpha$ indicates sampling of the ice or liquid phase. The
derivative of the melting temperature with respect to $\lambda$ is
then
\begin{equation}
  \label{eqn:dTdlam}
  \frac{\mrm{d}T_{\rm m}^{(r_{{\rm c},\lambda})}}{\mrm{d}\lambda} =
  \frac{T\big(x^{(\lambda)}_{\rm ice}-x^{(\lambda)}_{\rm liq}\big)}{h^{(\lambda)}_{\rm ice}-h^{(\lambda)}_{\rm liq}},
\end{equation}
where $h^{(\lambda)}_{\rm ice}$ and $h^{(\lambda)}_{\rm liq}$ are the
enthalpies per particle of ice and liquid, respectively, obtained from
trajectories using $U_\lambda$. Starting from $T_{\rm m}^{(8.5)}$,
$T_{\rm m}^{(9.25)}$ was obtained by integrating Eq.~\ref{eqn:dTdlam}
by fourth-order Runge-Kutta integration. This was then repeated,
starting from $T_{\rm m}^{(9.25)}$, to obtain an estimate for $T_{\rm
  m}^{(10.0)}$. We implemented $U_\lambda$ by tabulating the potential
at 0.0005\,\AA{} intervals for $1.8\,{\rm\AA}<r<10.1\,{\rm\AA}$, but
otherwise, simulation settings were the same as those described in
Secs.~\ref{subsec:WorkflowIce:AvgCell}
and~\ref{subsec:WorkflowLiq:dens}. Simulations were 5\,ns, following
0.5\,ns equilibration.

\section{Workflow: Liquid-vapor simulations}

\begin{figure}[tb]
  \centering
  \includegraphics[width=8cm]{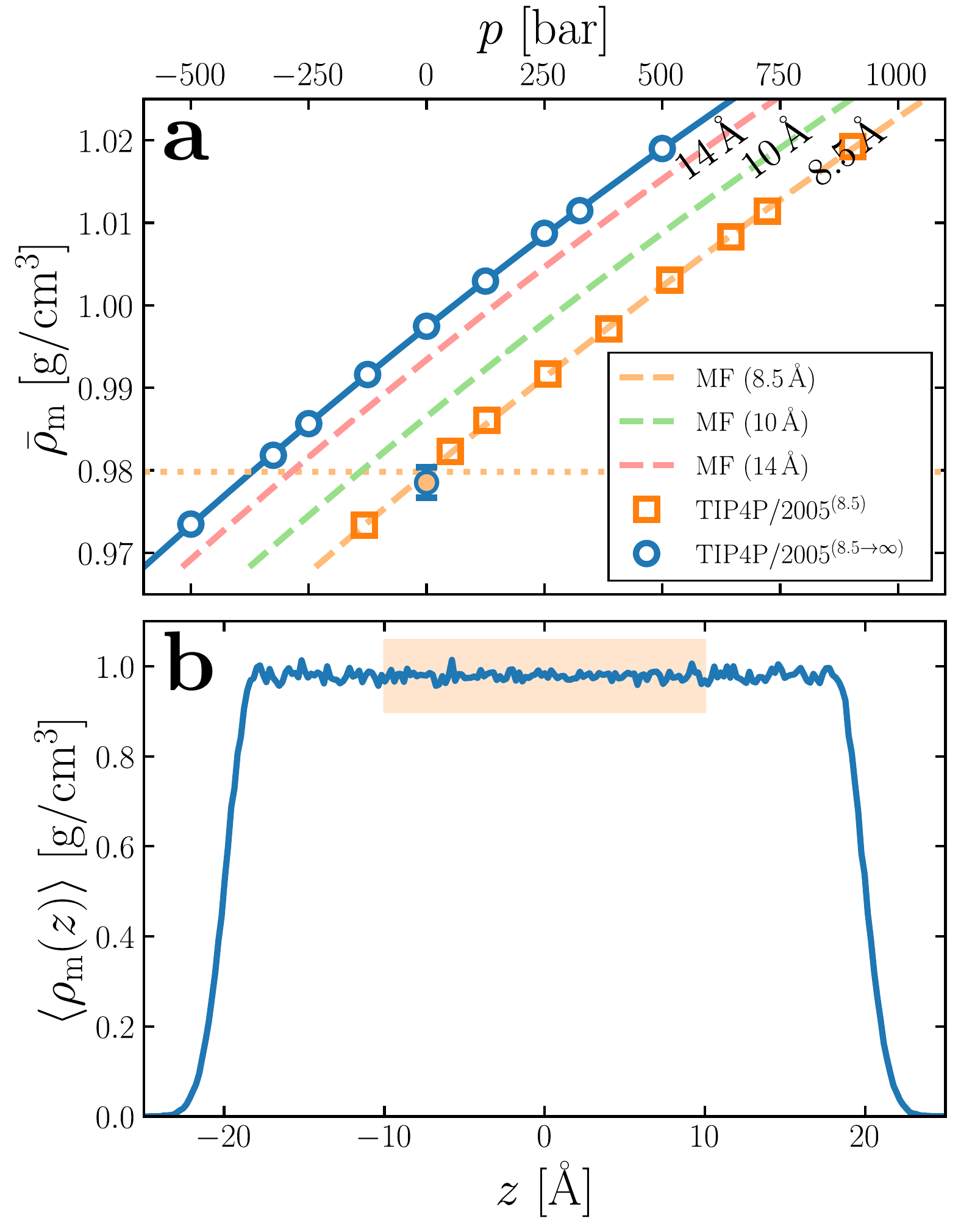}
  \caption{Evaluating the impact of $r_{\rm c}$ on $\bar{\rho}_{\rm
      m}$ for liquid TIP4P/2005 at 300\,K. (a) $\bar{\rho}_{\rm m}(p)$
    for a homogeneous system. White-filled circles show results from
    constant-$p$ simulations of TIP4P/2005$^{(8.5\to\infty)}$, and the
    solid blue line indicates a quadratic fit. Dashed lines indicate
    MF predictions (Eq.~\ref{eqn:Pdiff}) for different $r_{\rm c}$, as
    indicated in the legend. Orange squares show results from
    constant-$p$ simulations of TIP4P/2005$^{(8.5)}$. The dotted line
    indicates $\bar{\rho}_{\rm m}(0)$ for TIP4P/2005$^{(8.5)}$, which
    intercepts the TIP4P/2005$^{(8.5\to\infty)}$ results at $p\approx
    -370$\,bar. (b) $\langle\rho_{\rm m}(z)\rangle$ for a film of
    TIP4P/2005$^{(8.5\to\infty)}$ in contact with its vapor (only part
    of the simulation cell is shown). Spatially averaging
    $\langle\rho_{\rm m}(z)\rangle$ in the slab's interior, as
    indicated by the shaded region, gives an estimate $\bar{\rho}_{\rm
      m}(0)$, which is plotted with the orange-filled circle in (a).}
  \label{SMfig:RhoVsPress_300K}
\end{figure}

\noindent Simulations to produce Figs.~\ref{fig:RhoVsPress_300K}b
and~\ref{SMfig:RhoVsPress_300K}b comprised 512 water molecules, using
TIP4P/ice and TIP4P/2005, respectively. Simulation details are broadly
similar to those described in \ref{sec:WorkflowLiq}. The cross
sectional ($xy$) area of the simulation box was $19.7 \times
19.7$\,\AA$^2$, and its length normal ($z$) to the liquid-vapor
interface was 90\,\AA. To facilitate post-processing analysis,
repulsive walls as described in Ref.~\onlinecite{cox2020dielectric}
were placed at the edges of the simulation cell along $z$ to prevent
molecules escaping the primary simulation cell. The electric
displacement field along $z$ was set to zero, using the implementation
given in Refs.~\onlinecite{cox2019finite,sayer2019stabilization}; this
is formally equivalent to the commonly used slab correction of Yeh and
Berkowitz \cite{YehBerkowitz1999sjc}. Production simulations were
performed for 20\,ns following at least 0.5\,ns equilibration. A
Nos\'{e}-Hoover chain thermostat was used to maintain the temperature
at 300\,K. `Tail corrections' were formally applied, but as discussed
in the main text, this produces the same dynamics as the
`cut-and-shift' potential.

\section{Results for TIP4P/2005}
\label{sec:tip4p2005}

\noindent In this section, we present results obtained with
TIP4P/2005. While quantitative differences are expected, and indeed
observed, our general conclusions are unaffected by the choice of
water model. At $p=0$\,bar, we find $T_{\rm
  m}^{(8.5\to\infty)}=251.9$\,K in good agreement with $T_{\rm m}=252
\pm 6$\,K reported previously for $p=1$\,bar. We also see a modest
increase in melting temperature when using TIP4P/2005$^{(8.5)}$, with
$T_{\rm m}^{(8.5)} = 253.4$\,K and $T_{\rm m}^{({\rm MF,} 8.5)} =
254.0$\,K. The predictions of the mean-field prediction are supported
by Hamiltonian Gibbs-Duhem integration. Note that, unlike the results
for TIP4P/ice$^{(r_{\rm c})}$ reported in the main paper
\tcr{(Fig. 5b)}, the Hamiltonian Gibbs-Duhem simulations performed for
TIP4P/2005$^{(r_{\rm c})}$ were initiated from $T_{\rm m}^{({\rm MF,}
  8.5)}$ instead of $T_{\rm m}^{(8.5)}$ (indicated by the blue star in
Fig.~\ref{SMfig:BetaMuVsTemp_MF}).

\begin{figure}[tb]
  \centering
  \includegraphics[width=8cm]{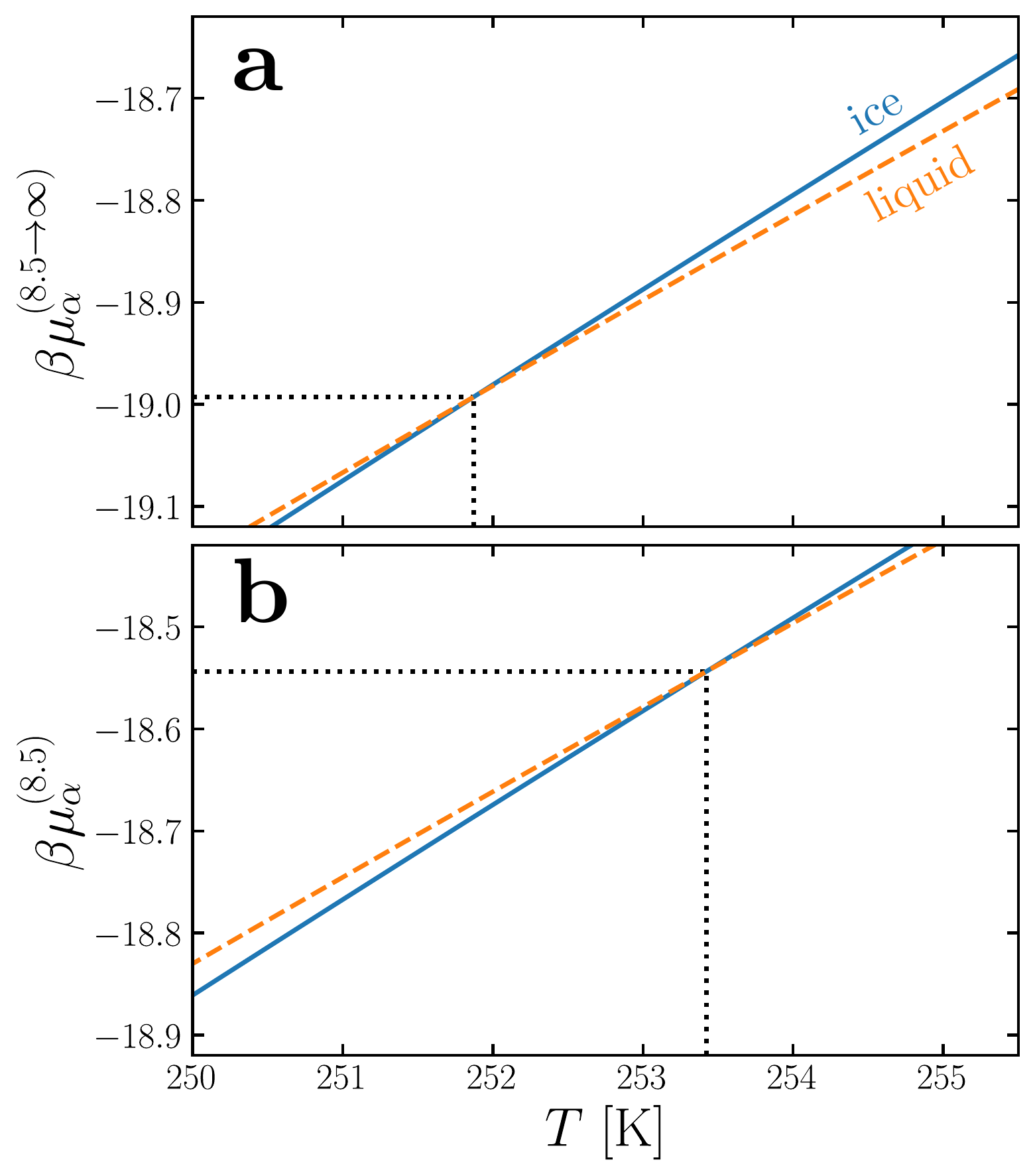}
  \caption{$\beta\mu_\alpha(T)$ at 0\,bar, with $\alpha = \text{`ice'
      or `liq'}$, for (a) TIP4P/2005$^{(8.5\to\infty)}$ and (b)
    TIP4P/2005$^{(8.5)}$. $T_{\rm m}$ is determined from the point of
    interception, as indicated by the black dotted lines, with
    $T^{(8.5\to\infty)}_{\rm m} = 251.9$\,K and $T^{(8.5)}_{\rm
      m}=253.4$\,K.}
  \label{SMfig:BetaMuVsTemp_TailShift}
\end{figure}

\begin{figure}[tb]
  \centering
  \includegraphics[width=8cm]{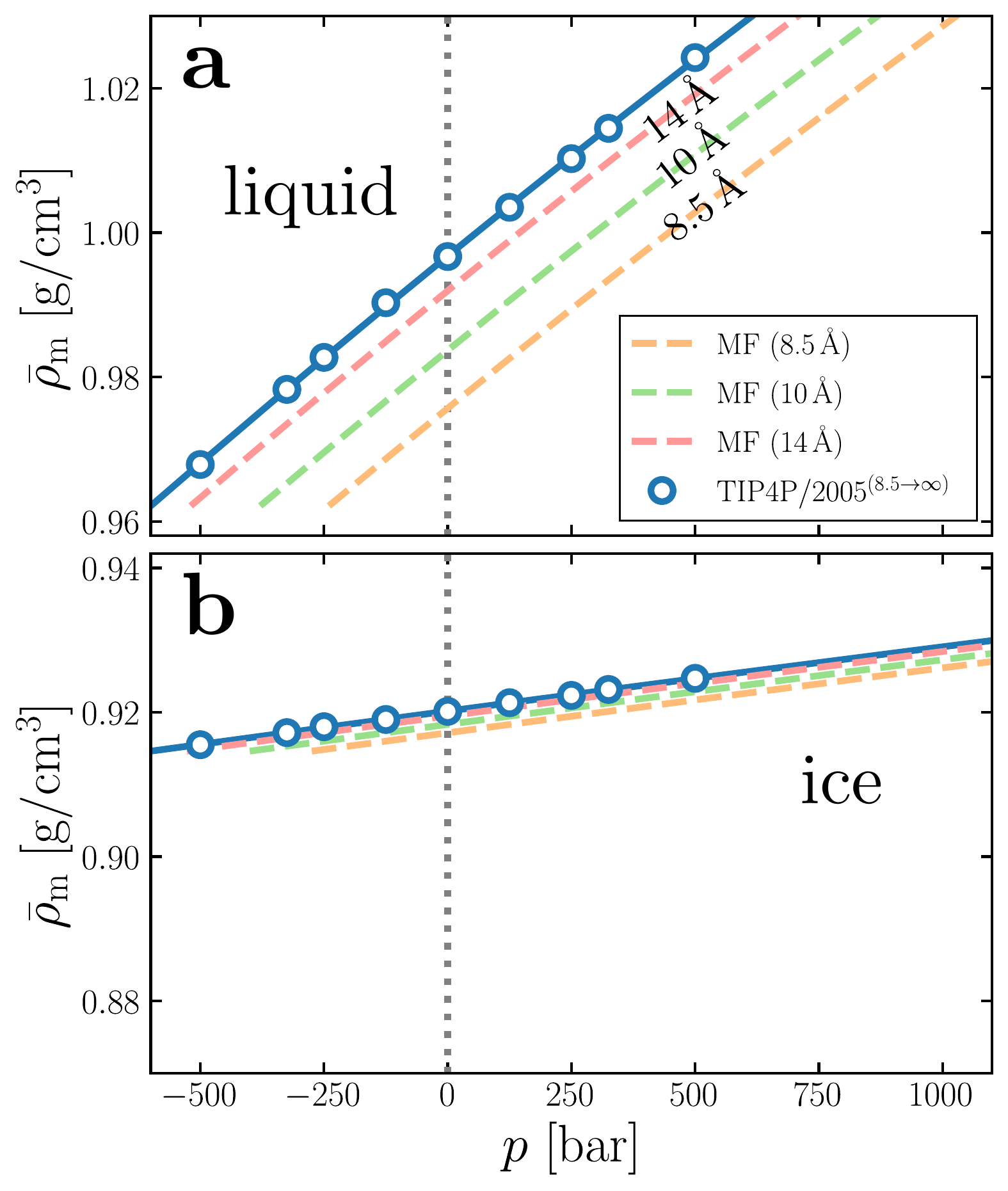}
  \caption{$\bar{\rho}_{\rm m}(p)$ at 252\,K for (a) liquid water and
    (b) ice. White-filled circles show results from constant-$p$
    simulations of TIP4P/2005$^{(8.5\to\infty)}$, and the solid blue
    line indicates a quadratic fit. Dashed lines indicate MF
    predictions (Eq.~\ref{eqn:Pdiff}) for different $r_{\rm c}$, as
    indicated in the legend, which are used to predict
    $\bar{\rho}_{\rm m}(0)$ for a given $r_{\rm c}$, i.e., where the
    dashed lines intersect the vertical gray dotted line.}
  \label{SMfig:RhoVsPress_Tm}
\end{figure}

\begin{figure}[tb]
  \centering
  \includegraphics[width=8cm]{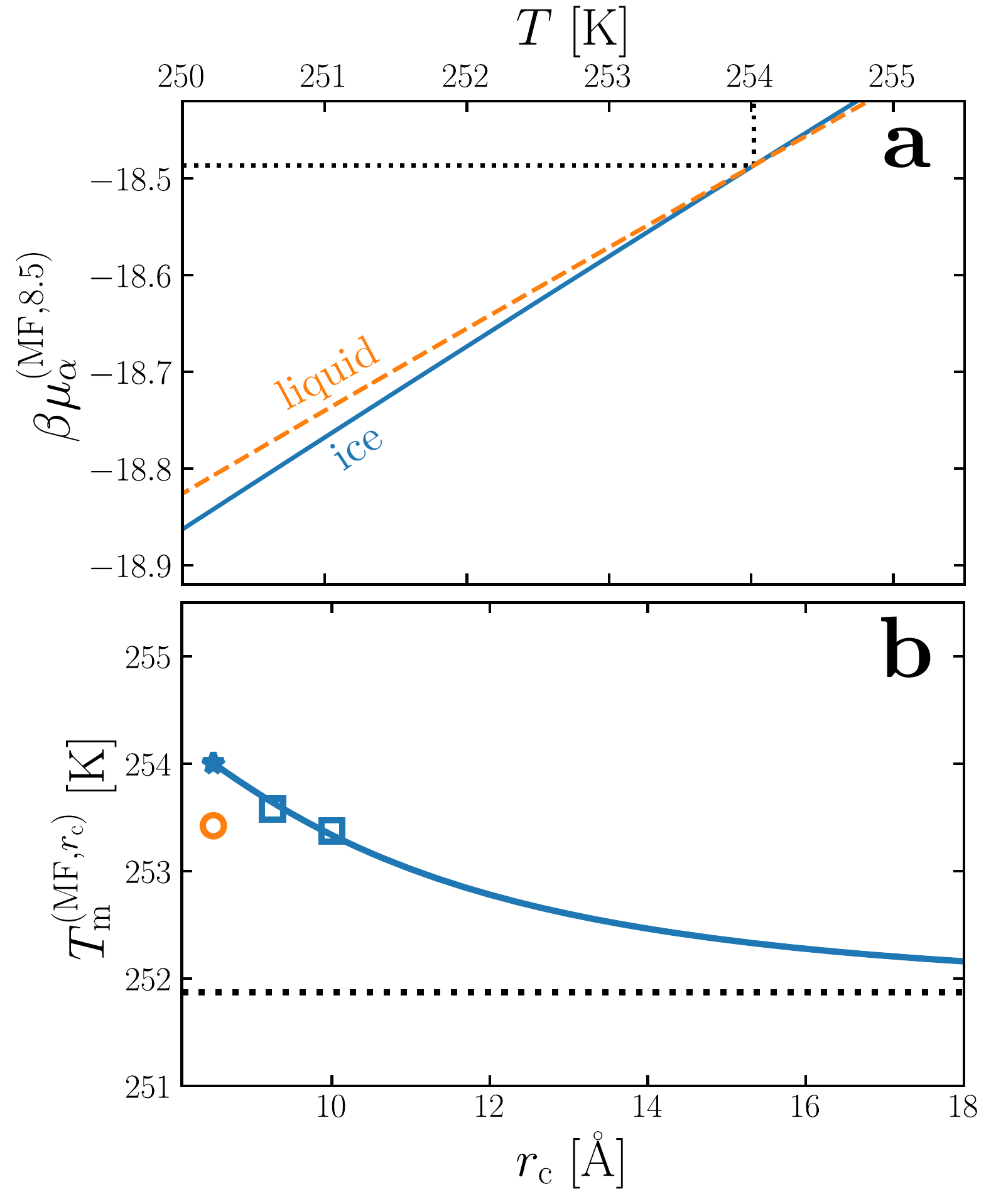}
  \caption{Predicting the effect of $r_{\rm c}$ on the melting
    temperature of TIP4P/2005 with MF theory. (a)
    $\beta\mu_\alpha^{({\rm MF,} 8.5)}(T)$ at $p=0$\,bar, with $\alpha
    = \text{`ice' or `liq'}$, obtained from
    Eq.~\ref{eqn:mudiff}. $T_{\rm m}^{({\rm MF,}8.5)}=254.0$\,K is
    determined from the point of interception, as indicated by the
    black dotted lines. (b) $T_{\rm m}^{({\rm MF,}r_{\rm c})}$ is
    shown by the solid blue line. The orange circle indicates $T_{\rm
      m}^{(8.5)}$ obtained from the free energy calculations described
    in Sec.~\ref{sec:meltingEinstein}, and the blue squares indicate
    $T_{\rm m}^{(9.25)}$ and $T_{\rm m}^{(10.0)}$ obtained from
    Hamiltonian Gibbs-Duhem integration, starting from $T_{\rm
      m}^{({\rm MF,}8.5)}$, which is marked with the blue star.}
  \label{SMfig:BetaMuVsTemp_MF}
\end{figure}

\section{Fitting coefficients}

\noindent In this section, we report the coefficients for the
quadratic polynomial $r_2p^2 + r_1p + r_0$ obtained using
\texttt{numpy}'s \texttt{polyfit} routine \cite{harris2020array}, as
shown in Figs.~\ref{fig:RhoVsPress_300K} and~\ref{fig:RhoVsPress_Tm}
in the main article, and Figs.~\ref{SMfig:RhoVsPress_300K}
and~\ref{SMfig:RhoVsPress_Tm}.

\subsection{Results for TIP4P/ice}

\begin{itemize}
\item{Liquid, 300\,K (Fig.~\ref{fig:RhoVsPress_300K}a):}
  \begin{itemize}
  \item[]{$r_2 = -3.822428\times 10^{-9}$\,g/(bar$^2$\,cm$^3$);}
  \item[]{$r_1 = 4.460206\times 10^{-5}$\,g/(bar\,cm$^3$);}
  \item[]{$r_0=9.939254\times 10^{-1}$\,g/cm$^{3}$.}
  \end{itemize}
\item{Liquid, 272\,K (Fig.~\ref{fig:RhoVsPress_Tm}a):}
  \begin{itemize}
  \item[]{$r_2 = -4.700729 \times 10^{-9}$\,g/(bar$^2\,$cm$^3$);}
  \item[]{$r_1 = 5.129812\times 10^{-5}$\,g/(bar\,cm$^3$);}
  \item[]{$r_0 = 9.898901\times 10^{-1}$\,g/cm$^{3}$.}
  \end{itemize}
\item{Ice, 272\,K (Fig.~\ref{fig:RhoVsPress_Tm}b):}
  \begin{itemize}
  \item[]{$r_2 = 1.442316 \times 10^{-11}$\,g/(bar$^2\,$cm$^3$);}
  \item[]{$r_1 = 8.358199\times 10^{-6}$\,g/(bar\,cm$^3$);}
  \item[]{$r_0 = 9.056778\times 10^{-1}$\,g/cm$^{3}$.}
  \end{itemize}
\end{itemize}

\newpage

\subsection{Results for TIP4P/2005}

\begin{itemize}
\item{Liquid, 300\,K (Fig.~\ref{SMfig:RhoVsPress_300K}a):}
  \begin{itemize}
  \item[]{$r_2 = -4.748523\times 10^{-9}$\,g/(bar$^2$\,cm$^3$);}
  \item[]{$r_1 = 4.561509\times  10^{-5}$\,g/(bar\,cm$^3$);}
  \item[]{$r_0=9.973669\times 10^{-1}$\,g/cm$^{3}$.}
  \end{itemize}
\item{Liquid, 252\,K (Fig.~\ref{SMfig:RhoVsPress_Tm}a):}
  \begin{itemize}
    \item[]{$r_2 = -3.051495\times 10^{-9}$\,g/(bar$^2\,$cm$^3$);}
    \item[]{$r_1 = 5.587068\times 10^{-5}$\,g/(bar\,cm$^3$);}
    \item[]{$r_0 = 9.967701\times 10^{-1}$\,g/cm$^{3}$.}
  \end{itemize}
\item{Ice, 252\,K (Fig.~\ref{SMfig:RhoVsPress_Tm}b):}
  \begin{itemize}
  \item[]{$r_2 = -2.226744\times 10^{-10}$\,g/(bar$^2\,$cm$^3$);}
  \item[]{$r_1 = 9.123390\times 10^{-6}$\,g/(bar\,cm$^3$);}
  \item[]{$r_0 = 9.201753\times 10^{-1}$\,g/cm$^{3}$.}
  \end{itemize}
\end{itemize}

\tcb{\section{Comment on the apparent role of impulsive forces}}

We have remarked in the main article that in the canoncial ensemble,
dynamics are unaffected by the choice of $U^{(r_{\rm c})}$
vs. $U^{(r_{\rm c}\to\infty)}$. While we have verified this directly
by comparing trajectories, and by checking the forces between a pair
of LJ particles (as implemented in \texttt{LAMMPS}), the form of
$u^{(r_{\rm c}\to\infty)}_{\rm LJ}$ given by \tcr{Eq.~3} suggests the
presence of an impulsive force at $r=r_{\rm c}$. Here will we
demonstrate that including impulsive forces would be inconsistent with
standard implementations of tail corrections.

Let us introduce a system with the following potential energy:
\begin{equation}
  \label{eqn:Utot-imp}
  U^{(r_{\rm c}!)}(\mbf{R}^N)
  = \sum_{i<j}^Nu^{(r_{\rm c}!)}_{\rm LJ}(|\mbf{r}^{\rm (O)}_{ij}|) + U_{\rm elec}(\mbf{R}^N),
\end{equation}
with
\begin{equation}
  \label{eqn:uLJimp}
  u^{(r_{\rm c}!)}_{\rm LJ}(r) = u^{(\infty)}_{\rm LJ}(r)h(r_{\rm c}-r),
\end{equation}
where $h(r)$ is the Heaviside step function. The potential energy
function $U^{(r_{\rm c}!)}$ describes a system where LJ interactions
are described by the unshifted LJ potential for $r\le r_{\rm c}$, and
abruptly vanish for $r>r_{\rm c}$. Forces due to the LJ interactions
are obtained by differentiation,
\begin{equation}
  f^{(r_{\rm c}!)}_{\rm LJ}(r) = f^{(\infty)}_{\rm LJ}(r)h(r_{\rm c}-r) + u^{(\infty)}_{\rm LJ}(r)\delta(r_{\rm c}-r).
\end{equation}
We clearly see an impulsive force at $r=r_{\rm c}$. Now consider the
average virial pressure:
\begin{align}
  p^{(r_{\rm c}!)} &= \frac{2\pi\bar{\rho}^2}{3}\int_0^{r_{\rm c}}\!\mrm{d}r\,r^3f^{(\infty)}_{\rm LJ}(r)g(r)
  + \frac{2\pi\bar{\rho}^2}{3}r_{\rm c}^3u^{(\infty)}_{\rm LJ}(r_{\rm c}), \\
  &= \frac{2\pi\bar{\rho}^2}{3}\int_0^{r_{\rm c}}\!\mrm{d}r\,r^3f^{(\infty)}_{\rm LJ}(r)g(r)
  + \frac{8\pi\epsilon\bar{\rho}^2\sigma^3}{3}\left[\bigg(\frac{\sigma}{r_{\rm c}}\bigg)^9-\bigg(\frac{\sigma}{r_{\rm c}}\bigg)^3\right],
  \label{eqn:VirialImp}
\end{align}
where we have assumed that $g_{\rm OO}(r\ge r_{\rm c})=1$. The second
term in Eq.~\ref{eqn:VirialImp}, which we will denote $\Delta
p^{(r_{\rm c}!)}$, is the impulsive contribution to the virial. For a
system where impulsive forces are present (whose dynamics in the $NVT$
ensemble in principle differ from $U^{(r_{\rm c})}$ and $U^{(r_{\rm
    c}\to\infty)}$ systems), one is required to add $\Delta p^{(r_{\rm
    c}!)}$ to the virial pressure, which in turn will affect the
dynamics in the $NpT$ ensemble. If we attempt to account for neglected
interactions beyond the cutoff in the usual fashion by simply adding the contribution
\begin{equation}
  \Delta_{\rm MF}p(r_{\rm c}) = \frac{2\pi\bar{\rho}^2}{3}\int_{r_{\rm c}}^\infty\!\mrm{d}r\,r^3f_{\rm LJ}^{(\infty)}(r) =
  \frac{32\pi\epsilon\bar{\rho}^2\sigma^3}{9}\Bigg[\bigg(\frac{\sigma}{r_{\rm c}}\bigg)^9-\frac{3}{2}\bigg(\frac{\sigma}{r_{\rm c}}\bigg)^3\Bigg]
\end{equation}
to $p^{(r_{\rm c}!)}$, we find an average virial pressure,
\begin{equation}
  \label{eqn:IncorrectVirial}
  \frac{2\pi\bar{\rho}^2}{3}\int_0^{r_{\rm c}}\!\mrm{d}r\,r^3f^{(\infty)}_{\rm LJ}(r)g(r) + \Delta p^{(r_{\rm c}!)}
+ \Delta_{\rm MF}p(r_{\rm c}),
\end{equation}
that does not approximately describe the average virial pressure of a
$U^{(\infty)}$ system.

Now consider a $U^{(r_{\rm c}\to\infty)}$ system. The LJ pair
potential is
\begin{equation}
  \label{eqn:uLJtc}
  u^{(r_{\rm c}\to\infty)}_{\rm LJ}(r) = u^{(\infty)}_{\rm LJ}(r)h(r_{\rm c}-r) + u^{(\infty)}_{\rm LJ}(r)h(r-r_{\rm c}),
\end{equation}
with the proviso that interactions for $r>r_{\rm c}$ are evaluated in
a mean field fashion. The forces are:
\begin{align}
  \label{eqn:fLJtc}
  f^{(r_{\rm c}\to\infty)}_{\rm LJ}(r) &= f^{(\infty)}_{\rm LJ}(r)h(r_{\rm c}-r)
  + f^{(\infty)}_{\rm LJ}(r)h(r-r_{\rm c}) + u^{(\infty)}_{\rm LJ}(r)\delta(r_{\rm c}-r)
  - u^{(\infty)}_{\rm LJ}(r)\delta(r-r_{\rm c}).
\end{align}
The impulsive forces at $r=r_{\rm c}$ cancel. Again, we consider the
average virial pressure:
\begin{align}
  p^{(r_{\rm c}\to\infty)} &= \frac{2\pi\bar{\rho}^2}{3}\int_0^{r_{\rm c}}\!\mrm{d}r\,r^3f^{(\infty)}_{\rm LJ}(r)g(r)
  + \Delta p^{(r_{\rm c}!)} - \Delta p^{(r_{\rm c}!)} + \Delta_{\rm MF}p(r_{\rm c}), \\
  &= \frac{2\pi\bar{\rho}^2}{3}\int_0^{r_{\rm c}}\!\mrm{d}r\,r^3f^{(\infty)}_{\rm LJ}(r)g(r)
  + \Delta_{\rm MF}p(r_{\rm c}). \label{eqn:VirialTC}
\end{align}
Equation~\ref{eqn:VirialTC} demonstrates that the standard `tail
correction,' $\Delta_{\rm MF}p$, is appropriate for a system that
employs $u^{(r_{\rm c}\to\infty)}_{\rm LJ}(r)$ \tcr{(Eq.~3)} to
describe explicit LJ interactions for $r\le r_{\rm c}$ in which the
apparent impulsive force at $r = r_{\rm c}$ is not included. In this
case, dynamics in the $U^{(r_{\rm c}\to\infty)}$ and $U^{(r_{\rm c})}$
systems are identical in the $NVT$ ensemble. It would be inconsistent
to use $\Delta_{\rm MF} p(r_{\rm c})$ in combination with a system
whose dynamics includes impulsive forces (see
\ref{eqn:IncorrectVirial}).

\end{document}